\documentclass[ALICE,manyauthors]{cernphprep}

\usepackage[comma,square,numbers,sort&compress]{natbib}
\usepackage{hyperref}
\usepackage{lineno}
\usepackage{color}
\usepackage{caption}
\usepackage{multirow}
\makeatletter{}

\newcommand{\energydensity}{\energydensityB}
\newcommand{\energydensityfactor}{\energydensityfactorB}

\newcommand{\energydensityB}{12.3 $\pm$ 1.0 GeV/fm$^3$\xspace}
\newcommand{\energydensitytauB}{12.3 $\pm$ 1.0 GeV/fm$^2$/$c$\xspace}
\newcommand{\energydensityfactorB}{2.3\xspace}
\newcommand{\RB}{7.17 fm\xspace}

\newcommand{\energydensitytauE}{12.5 $\pm$ 1.0 GeV/fm$^2$/$c$\xspace}

\newcommand{\centralet}{1737~$\pm$~6(stat.)~$\pm$~97~(sys.) GeV\xspace}

\newcommand{\centraletnpart}{9.02~$\pm$~0.03(stat.)~$\pm$~0.50~(sys.) GeV\xspace}
\newcommand{\ET}{$E_{\mathrm T}$\xspace}
\newcommand{\EThad}{$E_{\mathrm T}^{\mathrm{had}}$\xspace}
\newcommand{\ETpikp}{$E_{\mathrm T}^{\pi\mathrm{,K,p}}$\xspace}
\newcommand{\ETem}{$E_{\mathrm T}^{\mathrm{em}}$\xspace}

\newcommand{\ETbold}{$\mathbf{{\mathit E}_{\mathrm T}}$\xspace}
\newcommand{\EThadbold}{$\mathbf{{\mathit E}_{\mathrm T}^{\mathrm{had}}}$\xspace}
\newcommand{\ETembold}{$\mathbf{{\mathit E}_{\mathrm T}^{\mathrm{em}}}$\xspace}

\newcommand{\pT}{\ensuremath{p_{\rm T}}\xspace}
\newcommand{\pTbold}{\ensuremath{\mathbf{p_{\rm T}}}\xspace}

\newcommand{\Np}{$\langle N_{\mathrm{part}} \rangle$\xspace}
\newcommand{\Npart}{\Np}
\newcommand{\Nq}{$\langle N_{\mathrm{quark}} \rangle$\xspace}

\newcommand{\Nch}{$\langle N_{\mathrm{ch}} \rangle$\xspace}
\newcommand{\dEdx}{d$E$/d$x$\xspace}
\newcommand{\meandEtdeta}{$\langle $d$E_{\mathrm T}/$d$\eta \rangle$\xspace}
\newcommand{\meandEtdy}{$\langle $d$E_{\mathrm T}/$dy$ \rangle$\xspace}

\newcommand{\meanpTmeandNchdeta}{$\langle p_{\mathrm{T}} \rangle   \langle $d$N_{\mathrm{ch}} /$d$\eta \rangle$\xspace}
\newcommand{\meandEthaddetaPerNp}{$\langle $d$E_{\mathrm T}^{\mathrm{had}}/$d$\eta \rangle / \langle N_{\mathrm{part}}/2\rangle$\xspace}
\newcommand{\meandEtemdetaPerNp}{$\langle $d$E_{\mathrm T}^{\mathrm{em}}/$d$\eta \rangle / \langle N_{\mathrm{part}}/2\rangle$\xspace}
\newcommand{\meandEtdetaPerNp}{$\langle $d$E_{\mathrm T}/$d$\eta \rangle / \langle N_{\mathrm{part}}/2\rangle$\xspace}
\newcommand{\meandEtdetaPerNq}{$\langle $d$E_{\mathrm T}/$d$\eta \rangle / \langle N_{\mathrm{quark}}/2\rangle$\xspace}

\newcommand{\meandEtdetaOverdNchdeta}{$\langle $d$E_{\mathrm T}/$d$\eta \rangle$/$\langle $d$N_{\mathrm{ch}}/$d$\eta \rangle$\xspace}
\newcommand{\meanpT}{$\langle p_{\mathrm{T}} \rangle$\xspace}
\newcommand{\MeV}{MeV/$c$\xspace}
\newcommand{\GeV}{GeV/$c$\xspace}

\newcommand{\Fref}[1]{Fig.~\ref{#1}}
\newcommand{\Figure}[1]{Figure~\ref{#1}}
\newcommand{\Eref}[1]{Eq.~\ref{#1}}
\newcommand{\Erefs}[2]{Eqs.~\ref{#1} and~\ref{#2}}

\newcommand{\Cref}[1]{Chapter~\ref{#1}}

\newcommand{\Tref}[1]{Tab.~\ref{#1}}

\newcommand{\pp}{pp\xspace}
\newcommand{\Pb}{Pb--Pb\xspace}
\newcommand{\Au}{Au--Au\xspace}

\newcommand{\sqrts}{$\sqrt{s}$\xspace}
\newcommand{\sNN}{${\sqrt{s_{\mathrm {NN}}}}$\xspace}

\newcommand{\pikp}{$\pi^{\pm}$, K$^{\pm}$, p, and $\overline{\mathrm p}$\xspace}
\newcommand{\pikorp}{$\pi^{\pm}$, K$^{\pm}$, p, or $\overline{\mathrm p}$\xspace}
\newcommand{\pikpn}{$\pi^{\pm}$, K$^{\pm}$, p, $\overline{\mathrm p}$, n, and $\overline{\mathrm n}$\xspace}
\newcommand{\vzeros}{$\Lambda$, $\overline{\Lambda}$, and K$^{\mathrm 0}_{\mathrm S}$\xspace}
\newcommand{\categoryC}{$\Lambda$, $\overline{\Lambda}$, K$^{\mathrm 0}_{\mathrm S}$, $\Sigma^+$, $\Sigma^-$, and $\Sigma^0$\xspace}

\newcommand{\kzeroshort}{$K^{\mathrm 0}_{\mathrm S}$\xspace}

\newcommand{\neutralsundet}{K$^{\mathrm 0}_{\mathrm L}$, n, and $\overline{\mathrm n}$\xspace}
\newcommand{\lam}{$\Lambda$\xspace}
\newcommand{\alam}{$\overline{\Lambda}$\xspace}

\newcommand{\allemparticles}{$\pi^{\mathrm 0}$, $\omega$, $\eta$, e$^{\pm}$, and $\gamma$\xspace}
\newcommand{\otheremparticles}{$\omega$, $\eta$, e$^{\pm}$, and $\gamma$\xspace}
\newcommand{\pizero}{$\pi^{\mathrm 0}$\xspace}

\newcommand{\facc}{$f_{\mathrm{acc}}$\xspace}
\newcommand{\faccbold}{$\mathbf{{\mathit f}_{\mathrm{acc}}}$\xspace}
\newcommand{\fptcut}{$f_{\mathrm{p}_\mathrm{Tcut}}$\xspace}
\newcommand{\fptcutbold}{$\mathbf{{\mathit f}_{\mathrm{p}_\mathrm{Tcut}}}$\xspace}

\newcommand{\fneutral}{$f_{\mathrm{neutral}}$\xspace}
\newcommand{\ftotal}{$f_{\mathrm{total}}$\xspace}
\newcommand{\ftotalpaperone}{\ftotal = 0.553 $\pm$ 0.010\xspace}
\newcommand{\fneutralbold}{$\mathbf{{\mathit f}_{\mathrm{neutral}}}$\xspace}
\newcommand{\ftotalbold}{$\mathbf{{\mathit f}_{\mathrm{total}}}$\xspace}
\newcommand{\fpi}{$f_{\pi}$\xspace}
\newcommand{\fK}{$f_{K}$\xspace}
\newcommand{\fp}{$f_{p}$\xspace}
\newcommand{\flambda}{$f_{\Lambda}$\xspace}
\newcommand{\fem}{$f_{\mathrm{em}}$\xspace}
\newcommand{\fbkgd}{$f_{\mathrm{bg}}(p_{\mathrm T})$\xspace}

\newcommand{\fbkgdbold}{$\mathbf{{\mathit f}_{\mathrm{bg}}(p_{\mathrm T})}$\xspace}

\newcommand{\fnotid}{$f_{\mathrm{notID}}$\xspace}
\newcommand{\fnotidbold}{$f_\mathbf{{\mathrm{notID}}}$\xspace}
\newcommand{\eff}{$\epsilon(p_{\mathrm T})$\xspace}
\newcommand{\effbold}{$\mathbf{\epsilon(p_{\mathrm T})}$\xspace}

\newcommand{\femin}{$f_{E_{\mathrm{Tmin}}}$\xspace}
\newcommand{\feminbold}{$\mathbf{f_{E_{\mathrm{Tmin}}}}$\xspace}
\newcommand{\fenl}{$f_{E_{\mathrm{NL}}}$\xspace}
\newcommand{\fenlbold}{$\mathbf{f_{E_{\mathrm{NL}}}}$\xspace}
\newcommand{\fereceff}{$\epsilon_{\gamma}$\xspace}
\newcommand{\fereceffbold}{$\mathbf{\epsilon_{\gamma}}$\xspace}

\newcommand{\fetbkgd}{$E_{\mathrm T}^{\mathrm {bkgd}}$\xspace}
\newcommand{\fetbkgdbold}{$\mathbf{E_{\mathrm T}^{\mathrm {bkgd}}}$\xspace}
\newcommand{\sims}{HIJING+GEANT simulations\xspace}

\newcommand{\fbkgdemet}{$f_{\mathrm {bkgd}}$\xspace}
\newcommand{\deltamatched}{$\delta_{\mathrm m}$\xspace}
\newcommand{\npart}{$\langle N_{\mathrm{part}} \rangle$\xspace}
\newcommand{\npartpair}{$\langle N_{\mathrm{part}}/2 \rangle$\xspace}
\newcommand{\nquark}{$\langle N_{\mathrm{quark}} \rangle$\xspace}

\usepackage{etoolbox}

\providetoggle{preprint}
\settoggle{preprint}{true}
\newlength{\figurewidth}
\iftoggle{preprint}
{\setlength{\figurewidth}{10cm} }{\setlength{\figurewidth}{8cm} }

\begin{document}
\begin{titlepage}
\PHyear{2016}
\PHnumber{071}      
\PHdate{11 March}  
\title{Measurement of transverse energy at midrapidity\\in Pb--Pb collisions at $\mathbf{\sqrt{{\textit s}_{\rm NN}}}$ = 2.76 TeV}
\ShortTitle{Transverse energy at midrapidity in Pb--Pb collisions}   
\Collaboration{ALICE Collaboration\thanks{See Appendix~\ref{app:collab} for the list of collaboration members}}
\ShortAuthor{ALICE Collaboration} 
\begin{abstract}
\makeatletter{}We report the transverse energy (\ET) measured with ALICE at midrapidity in \Pb collisions at \sNN = 2.76 TeV as a function of centrality.  
The transverse energy was measured using identified single particle tracks. The measurement was cross checked using the electromagnetic calorimeters and the transverse momentum distributions of identified particles previously reported by ALICE.
The results are compared to theoretical models as well as to results from other experiments.
The mean \ET per unit pseudorapidity ($\eta$), \meandEtdeta, in 0--5\% central collisions is \centralet.
We find a similar centrality dependence of the shape of \meandEtdeta as a function of the number of participating nucleons to that seen at
lower energies. The growth in \meandEtdeta at the LHC energies exceeds extrapolations of low energy data.
We observe a nearly linear scaling of \meandEtdeta with the number of quark participants. 
With the canonical assumption of a 1 fm/$c$ formation time, we estimate that the energy density in 0--5\% central \Pb collisions at \sNN = 2.76 TeV is \energydensity and that 
the energy density at the most central 80 fm$^2$ of the collision is at least 21.5 $\pm$ 1.7 GeV/fm$^3$.  This is roughly \energydensityfactor times that observed in 0--5\% central \Au collisions at \sNN = 200 GeV.
\end{abstract}
\end{titlepage}
\setcounter{page}{2}

\section{Introduction}\label{Sec:Introduction}
\makeatletter{} Quantum Chromodynamics (QCD) predicts a phase transition of nuclear matter to a plasma of quarks and gluons at energy densities above about 0.2-1 GeV/fm$^3$~\cite{Karch:2002abc,Bazavov:2014pvz}.  This matter, called Quark--Gluon Plasma (QGP), is produced in high energy nuclear collisions~\cite{Adcox:2004mh,Adams:2005dq,Back:2004je,Arsene:2004fa,Aamodt:2011mr,Aamodt:2010cz,Aamodt:2010jd,Aamodt:2010pb,Aamodt:2010pa,Aad:2010bu,Chatrchyan:2011sx,Chatrchyan:2011pb,Chatrchyan:2012ta,ATLAS:2011ag,ATLAS:2011ah} and its properties are being investigated at the Super Proton Synchrotron (SPS), the Relativistic Heavy Ion Collider (RHIC) and the Large Hadron Collider (LHC).  The highest energy densities are achieved at the LHC in \Pb collisions.  

The mean transverse energy per unit pseudorapidity \meandEtdeta conveys information about how much of the initial longitudinal energy carried by the incoming nuclei is converted into energy carried by the particles produced transverse to the beam axis.
The transverse energy at midrapidity is therefore a measure of the stopping power of nuclear matter.  
By using simple geometric considerations~\cite{Bjorken:1982qr} \meandEtdeta can provide information on the energy densities attained.  Studies of the centrality and \sNN dependence of \meandEtdeta therefore provide insight into the conditions prior to thermal and chemical equilibrium. 

The \meandEtdeta has been measured at the AGS by E802~\cite{Ahle:1994nk} and E814/E877~\cite{Barrette:1993pm}, at the SPS by NA34~\cite{Akesson:1987kh}, NA35~\cite{Bamberger:1986mt}, NA49~\cite{Margetis:1994tt}, and WA80/93/98~\cite{WA80Et:1991,Aggarwal:2000bc}, at RHIC by PHENIX~\cite{Adler:2004zn,Adcox:2001ry,Adler:2013aqf} and STAR~\cite{Adams:2004cb}, and at the LHC by CMS~\cite{CMSET:2012}, covering nearly three orders of magnitude of \sNN. The centrality dependence has also been studied extensively with \meandEtdeta at midrapidity scaling nearly linearly with the collision volume, or equivalently, the number of participating nucleons at lower energies~\cite{Bialas:1976ed,WA80Et:1991,Miller:2007}. Further studies of heavy-ion collisions revealed deviations from this simple participant scaling law~\cite{Aggarwal:2000bc}. The causes of this deviation from linearity are still actively discussed and might be related to effects from minijets~\cite{Wang:2000bf,Trainor:2015ida} or constituent quark scaling~\cite{Eremin:2003,Adler:2014a}.

The ALICE detector~\cite{Aamodt:2008zz} has precision tracking detectors and
electromagnetic calorimeters, enabling several different methods for measuring \ET.
In this paper we discuss measurements of \meandEtdeta in \Pb collisions at \sNN = 2.76 TeV using the tracking detectors alone and using the combined information from the tracking detectors and the electromagnetic calorimeters.  In addition we compare to calculations of \meandEtdeta from the measured identified particle transverse momentum distributions. 
Measurements from the tracking detectors alone provide the highest precision.  We compare our results to theoretical calculations and measurements at lower energies. 
\section{Experiment}
\makeatletter{}A comprehensive description of the ALICE detector can be found in~\cite{Aamodt:2008zz}.
This analysis uses the V0, Zero Degree Calorimeters (ZDCs), the Inner Tracking System (ITS), the Time Projection
Chamber (TPC), the ElectroMagnetic Calorimeter (EMCal), and the PHOton Spectrometer (PHOS), all of which are located inside a 0.5 T solenoidal magnetic field. 
The V0 detector~\cite{Abbas:2013taa} consists of two scintillator hodoscopes covering the pseudorapidity ranges $-$3.7 $< \eta <$ $-$1.7 and 2.8 $< \eta <$ 5.1.  The ZDCs each consist of a neutron calorimeter between the beam pipes downstream of the dipole magnet and a proton calorimeter external to the outgoing beam pipe.

The TPC~\cite{Alme:2010ke}, the main tracking detector at midrapidity, is a cylindrical drift detector filled with a Ne--CO$_2$
gas mixture.  The active volume is nearly 90 m$^3$ and has inner and outer radii of 0.848~m and 2.466~m, respectively.  
It provides particle identification via the measurement of the specific ionization energy loss (\dEdx) with a resolution of 
~5.2\% and ~6.5\%  in peripheral and central collisions, respectively.

The ITS~\cite{Aamodt:2008zz} consists of the Silicon Pixel Detector with layers at radii of 3.9~cm and 7.6~cm, the Silicon Drift Detector with layers at radii of 15.0~cm and 23.9~cm, and the Silicon Strip Detector with layers at radii of 38.0 and 43.0~cm.
The TPC and ITS are aligned to within a few hundred $\mu$m using cosmic ray and \pp collision data~\cite{Aamodt:2010aa}.

The EMCal~\cite{Allen:2009aa,Cortese:2008zza} is a lead/scintillator sampling calorimeter covering $|\eta| <$ 0.7 in pseudorapidity and 100$^{\circ}$ in azimuth in 2011.  The EMCal consists of 11520 towers, each with transverse size 6 cm  $\times$ 6 cm, or approximately twice the effective Moli\`ere radius.  
The relative energy resolution is $\sqrt{0.11^2/E+0.017^2}$ where the energy $E$ is measured in GeV~\cite{Allen:2009aa}.
Clusters are formed by combining signals from adjacent towers.  Each cluster is required to have only one local energy maximum.  Noise is suppressed by requiring a minimum tower energy of 0.05~GeV.
For this analysis we use clusters within $|\eta| <$ 0.6.  The PHOS~\cite{Dellacasa:1999kd} is a lead tungstate calorimeter covering  $|\eta| <$ 0.12 in pseudorapidity and 60$^{\circ}$ in azimuth.  The PHOS consists of three modules of 64 $\times$ 56  towers each, with each tower having a transverse size of 2.2 cm $\times$ 2.2 cm, comparable to the Moli\`ere radius.  The relative energy resolution is $\sqrt{0.013^2/E^2 + 0.036^2/E + 0.01^2}$ where the energy $E$ is measured in GeV~\cite{Aleksandrov:2005yu}.

The minimum-bias trigger for \Pb collisions in 2010 was defined by a combination of hits in the V0 detector and the two innermost (pixel) layers of the ITS~\cite{Aamodt:2010cz}.  In 2011 the minimum-bias trigger signals in both neutron ZDCs were also required~\cite{Abelev:2014ffa}.
The collision centrality is determined by comparing the multiplicity measured in the V0 detector to Glauber model simulations of the multiplicity~\cite{Aamodt:2010cz,Abbas:2013taa}.  These calculations are also used to determine the number of participating nucleons, \npart.  We restrict our analysis to the 0--80\% most central collisions.  For these centralities corrections due to electromagnetic interactions and trigger inefficiencies are negligible.  We use data from approximately 70k 0-80\% central events taken in 2011 for the tracking detector and EMCal measurements and data from approximately 600k 0-80\% central events taken in 2010 for the PHOS measurement.  We focus on a small event sample where the detector performance was uniform in order to simplify efficiency corrections since the measurement is dominated by systematic uncertainties.

Tracks are reconstructed using both the TPC and the ITS.  Tracks are selected by requiring  that they cross at least 70 rows and requiring a $\chi^2$ per space point $<$ 4. Tracks are restricted to $|\eta|<$ 0.6. Each track is required to have at least one hit in one of the two innermost ITS layers and a small distance of closest approach (DCA) to the primary vertex in the $xy$ plane as a function of transverse momentum (\pT), defined by \mbox{DCA$_{\mathrm {xy}}$ $<$ (0.0182+0.035/ $p_{\mathrm T}^{1.01}$)} cm where \pT is in \GeV. The distance of closest approach in the $z$ direction is restricted to DCA$_{\mathrm z}$ $<$ 2 cm.  This reduces the contribution from secondary particles from weak decays, which appear as a background.  With these selection criteria tracks with transverse momenta \pT $>$ 150 \MeV can be reconstructed.  The typical momentum resolution for low momentum tracks, which dominate \ET measurements, is $\Delta$\pT/\pT~$\approx$~1\%.  The reconstruction efficiency varies with \pT~and ranges from about 50\% to 75\%~\cite{Abelev:2014ffa}.

Particles are identified through their specific energy loss, \dEdx, in the TPC when possible.
The \dEdx is calculated using a truncated-mean procedure and compared to the \dEdx expected for a given particle species using a Bethe-Bloch parametrization. The deviation from the expected \dEdx value is expressed in units of the energy-loss resolution $\sigma$~\cite{Abelev:2013vea}. 
Tracks are identified as arising from a kaon if they are within 3$\sigma$ from the expected \dEdx for a kaon, more than 3$\sigma$ from the expected \dEdx for a proton or a pion, and have \pT $<$ 0.45 \GeV.  Tracks are identified as arising from (anti)protons if they are within 3$\sigma$ from the expected \dEdx for (anti)protons, more than 3$\sigma$ from the expected \dEdx for kaons or pions, and have \pT $<$ 0.9 \GeV.  Tracks are identified as arising from an electron (positron) and therefore excluded from the measurement of \ETpikp if they are within 2$\sigma$ from the expected \dEdx for an electron (positron), more than 4$\sigma$ from the expected \dEdx for a pion, and more than 3$\sigma$ from the expected \dEdx for a proton or kaon.  With this algorithm approximately 0.1\% of tracks arise from electrons or positrons misidentified as arising from pions and fewer than 0.1\% of tracks are misidentified as arising from kaons or protons.  Any track not identified as a kaon or proton is assumed to arise from a pion and the measurement must be corrected for the error in this assumption.  

The PHOS and EMCal are used to measure the electromagnetic energy component of the \ET and to demonstrate consistency between methods.  Data from 2011 were used for the EMCal analysis due to the larger EMCal acceptance in 2011.  Data from one run in 2010 were used for the PHOS due to better detector performance and understanding of the
calibrations in that run period.  The EMCal has a larger acceptance, but the PHOS has a better energy resolution.  There is also a lower material budget in front of the PHOS than the EMCal.  This provides an additional check on the accuracy of the measurement.
 
\section{Method}\label{Sec:Methods}
\makeatletter{}
Historically most \ET measurements have been performed using calorimeters, and the commonly accepted operational definition of \ET is therefore based on the energy $E_j$ measured in the calorimeter's $j$′th tower
\begin{equation}\label{Eq:ET_Operational_Definition}
       E_{\mathrm T} = \sum_{{\mathrm j}=1}^{\mathrm M}E_{\mathrm j} \sin\theta_{\mathrm j}
\end{equation} 
\noindent where j runs over all M towers in the calorimeter and $\theta_{\mathrm j}$ is the polar angle of the calorimeter tower. 
The transverse energy can also be calculated using single particle tracks. In that case, the index, j,  in \Eref{Eq:ET_Operational_Definition} runs over the M measured particles instead of calorimeter towers, and $\theta_{\mathrm j}$ is the particle emission angle. In order to be compatible with the \ET of a calorimetry measurement, the energy $E_{\mathrm j}$ of \Eref{Eq:ET_Operational_Definition} must be replaced with the single particle energies
\begin{equation}\label{Eq:ET_Calorimeter_Response}
       E_{\mathrm j} = \left\{ 
	\begin{array}{ll}
                       E_{\mathrm {kin}} & \mbox{for baryons} \\
                       E_{\mathrm {kin}} +2mc^2  & \mbox{for anti-baryons} \\
                       E_{\mathrm {kin}} + mc^2 & \mbox{for all other particles.} 
	\end{array}
	\right.
\end{equation} 
This definition of \ET was used in the measurements of the transverse energy by CMS~\cite{CMSET:2012} (based on calorimetry), PHENIX~\cite{Adler:2004zn} (based on electromagnetic calorimetry), and STAR~\cite{Adams:2004cb} (based on a combination of electromagnetic calorimetry and charged particle tracking). To facilitate comparison between the various data sets the definition of \ET given by \Erefs{Eq:ET_Operational_Definition}{Eq:ET_Calorimeter_Response} is used here.

It is useful to classify particles by how they interact with the detector.  We define the following categories of final state particles:
\begin{description}
 \item[A] \pikp: Charged particles measured with high efficiency by tracking detectors
 \item[B] \allemparticles: Particles measured with high efficiency by electromagnetic calorimeters
 \item[C] \categoryC: Particles measured with low efficiency in tracking detectors and electromagnetic calorimeters
 \item[D] \neutralsundet: Neutral particles not measured well by either tracking detectors or electromagnetic calorimeters.
\end{description}
\noindent  The total \ET is the sum of the \ET observed in final state particles in categories A-D.  Contributions from all other particles are negligible.  In HIJING 1.383~\cite{Wang:1991hta} simulations of \Pb collisions at \sNN = 2.76 TeV the next largest contributions come from the $\Xi$($\overline{\Xi}$) and $\Omega$($\overline{\Omega}$) baryons with a total contribution of about 0.4\% of the total \ET, much less than the systematic uncertainty on the final value of \ET.
The \ET from unstable particles with $c\tau<$ 1~cm is taken into account through the \ET from their decay particles.

When measuring \ET using tracking detectors, the primary measurement is of particles in category A and corrections must be applied to take into account the \ET which is not observed from particles in categories B-D.  
In the hybrid method the \ET from particles in category A is measured using tracking detectors and the \ET from particles in category B is measured by the electromagnetic calorimeter.
An electromagnetic calorimeter has the highest efficiency for measuring particles in category B, although there is a substantial background from particles in category A.  The \ET from categories C and D, which is not well measured by an electromagnetic calorimeter, must be corrected for on average. 
Following the convention used by STAR, we define \EThad to be the \ET measured from particles in category A and scaled up to include particles in categories C and D and \ETem to be the \ET measured in category B.  The total \ET is given by
\begin{equation}\label{Eq:ETtotaleq}
 E_{\mathrm T} = E_{\mathrm T}^{\mathrm {had}} + E_{\mathrm T}^{\mathrm {em}}.
\end{equation} 
\noindent We refer to \EThad as the hadronic \ET and \ETem as the electromagnetic \ET.  We note that \EThad and \ETem are operational definitions based on the best way to observe the energy deposited in various detectors and that the distinction is not theoretically meaningful.

Several corrections are calculated using HIJING~\cite{Wang:1991hta} simulations.  The propagation of final state particles in these simulations through the ALICE detector material is described using GEANT 3~\cite{Brun:1994aa}.  Throughout the paper these are described as \sims.  

\makeatletter{}\subsection{Tracking detector measurements of \ETbold}\label{Sec:HadEt}
The measurements of the total \ET using the tracking detectors and of the hadronic \ET are closely correlated because the direct measurement in both cases is \ETpikp, the \ET from \pikp from the primary vertex.
All contributions from other categories are treated as background.  For \EThad the \ET from categories C and D is corrected for on average and for the total \ET the contribution from categories B, C, and D is corrected for on average.  Each of these contributions is taken into account with a correction factor.  

The relationship between the measured track momenta and \ETpikp is given by
\begin{equation}\label{Eq:pikpet}
 \frac{ {\mathrm d}E_{\mathrm T}^{\pi\mathrm{,K,p}}}{{\mathrm d}\eta} = 
 \frac{1}{\Delta\eta}
\frac{1}{f_{\mathrm{p_Tcut}}} 
\frac{1}{f_{\mathrm{notID}}} 
\sum\limits_{{\mathrm i}=1}^{\mathrm n}
\frac{f_{\mathrm{bg}}(p_{\mathrm T}^{\mathrm i})}{\epsilon(p_{\mathrm
T}^{\mathrm i})} E_{\mathrm i} \sin \theta_{\mathrm i}
\end{equation}
where i runs over the n reconstructed tracks and $\Delta\eta$ is the pseudorapidity range used in the analysis, \eff corrects for the finite track reconstruction efficiency and acceptance, \fbkgd corrects for the \vzeros daughters and electrons that pass the primary track quality cuts, \fnotid corrects for particles that could not be identified unambiguously through their specific energy loss \dEdx in the TPC, and \fptcut corrects for the finite detector acceptance at low momentum.
Hadronic \ET is given by \EThad = \ETpikp/\fneutral where \fneutral is the fraction of \EThad from \pikp
and total \ET is given by \ET = \ETpikp/\ftotal where \ftotal is the fraction of \ET from \pikp.
The determination of each of these corrections is given below and the systematic uncertainties are summarized in \Tref{Tab:hadCorrectionSysErrors}.  Systematic uncertainties are correlated point to point.

\subsubsection{Single track efficiency$\times$acceptance \effbold}
The single track efficiency$\times$acceptance is determined by comparing the primary yields to the reconstructed yields using \sims, as described in~\cite{Abelev:2014uua}.  When a particle can be identified as a \pikorp using the algorithm described above, the efficiency for that particle is used.  Otherwise the particle-averaged efficiency is used.  The 5\% systematic uncertainty is determined by the difference between the fraction of TPC standalone tracks matched with a hit in the ITS in simulations and data.  

\subsubsection{Background \fbkgdbold}
The background comes from photons which convert to $e^+e^-$ in the detector and decay daughters from \vzeros which are observed in the tracking detectors but do not originate from primary \pikp.  This is determined from \sims.  The systematic uncertainty on the background due to conversion electrons is determined by varying the material budget in the \sims by $\pm$10\% and found to be negligible compared to other systematic uncertainties.  The systematic uncertainty due to \vzeros daughters is sensitive to both the yield and the shape of the \vzeros spectra.  
To determine the contribution from \vzeros decay daughters and its systematic uncertainty the spectra in simulation are reweighted to match the data and the yields are varied within their uncertainties~\cite{Abelev:2013xaa}.  
Because the centrality dependence is less than the uncertainty due to other corrections, a constant correction of 0.982 $\pm$ 0.008 is applied across all centralities .

\subsubsection{Particle identification \fnotidbold}
The \ET of particles with 0.15 $<$ \pT $<$ 0.45 \GeV with a \dEdx within two standard deviations of the expected \dEdx for kaons is calculated using the kaon mass and the \ET of particles with 0.15 $<$ \pT $<$ 0.9 \GeV with a \dEdx within two standard deviations of the expected \dEdx for (anti)protons is calculated using the (anti)proton mass.  The \ET of all other particles is calculated using the pion mass.  Since the average transverse momentum is $\langle p_{\mathrm T}\rangle$ = 0.678 $\pm$ 0.007 \GeV for charged particles~\cite{Abelev:2013bla} and over 80\% of the particles created in the collision are pions~\cite{Abelev:2013vea}, most particles can be identified correctly using this algorithm.  At high momentum, the difference between the true \ET and the \ET calculated using the pion mass hypothesis for kaons and protons is less than at low \pT.  This is therefore a small correction.  Assuming that all kaons with 0.15 $<$ \pT $<$ 0.45 \GeV and (anti)protons with 0.15 $<$ \pT $<$ 0.9 \GeV are identified correctly and using the identifed \pikp spectra~\cite{Abelev:2013vea} gives \fnotid = 0.992 $\pm$ 0.002.  The systematic uncertainty is determined from the uncertainties on the yields.

Assuming that 5\% of kaons and protons identified using the particle identification algorithm described above are misidentified as pions only decreases \fnotid by 0.0002, less than the systematic uncertainty on \fnotid.  This indicates that this correction is robust to changes in the mean \dEdx expected for a given particle and its standard deviation.  
We note that either assuming no particle identification or doubling the number of kaons and protons only decreases \fnotid by 0.005.

\subsubsection{Low \pTbold\ acceptance \fptcutbold}
The lower momentum acceptance of the tracking detectors is primarily driven by the magnetic field and the inner radius of the active volume of the detector.  
Tracks can be reliably reconstructed in the TPC for particles with \pT $>$ 150 \MeV.  The fraction of \ET carried by particles below this momentum cut-off is determined by \sims.  To calculate the systematic uncertainty we follow the prescription given by STAR~\cite{Adams:2004cb}.  The fraction of \ET contained in particles below 150 \MeV is calculated assuming that all particles below this cut-off have a momentum of exactly 150 \MeV to determine an upper bound, assuming that they have a momentum of 0 \MeV to determine a lower bound, and using the average as the nominal value.  Using this prescription, \fptcut = 0.9710 $\pm$ 0.0058.  We note that \fptcut is the same within systematic uncertainties when calculated from PYTHIA simulations~\cite{Sjostrand:2006za} for \pp collisions with \sqrts = 0.9 and 8 TeV, indicating that this is a robust quantity.

\subsubsection{Correction factors \fneutralbold and \ftotalbold}\label{Sec:SpectraCalc}
Under the assumption that the different states within an isospin multiplet and particles and antiparticles have the same \ET, \fneutral can be written as
\begin{equation}\label{Eq:fneutral2}
 f_{\mathrm{neutral}} = \frac{  2E_{\mathrm T}^{\pi}+2E_{\mathrm T}^{\mathrm K}+2E_{\mathrm T}^{\mathrm p}  }{ 3E_{\mathrm T}^{\pi}+4E_{\mathrm T}^{\mathrm K}+4E_{\mathrm T}^{\mathrm p}+2E_{\mathrm T}^{\Lambda}+6E_{\mathrm T}^{\Sigma} }
\end{equation}
\noindent and \ftotal can be written as
\begin{equation}\label{Eq:ftotal2}
 f_{\mathrm{total}} = \frac{  2E_{\mathrm T}^{\pi}+2E_{\mathrm T}^{\mathrm K}+2E_{\mathrm T}^{\mathrm p}  }{ 3E_{\mathrm T}^{\pi}+4E_{\mathrm T}^{\mathrm K}+4E_{\mathrm T}^{\mathrm p}+2E_{\mathrm T}^{\Lambda}+6E_{\mathrm T}^{\Sigma}+ E_{\mathrm T}^{\omega,\eta,{\mathrm e}^{\pm},\gamma} }.
\end{equation}
\noindent  where $E_{\mathrm T}^{\mathrm K}$ is the \ET from one kaon species, $E_{\mathrm T}^{\pi}$ is the \ET from one pion species, $E_{\mathrm T}^{\mathrm p}$ is the average of the  \ET from protons and antiprotons, $E_{T}^{\Lambda}$ is the average \ET from $\Lambda$ and $\overline{\Lambda}$, and $E_{T}^{\Sigma}$ is the average \ET from $\Sigma^+$, $\Sigma^-$, and $\Sigma^0$ and their antiparticles.  
The contributions $E_{\mathrm T}^{\pi}$, $E_{\mathrm T}^{\mathrm K}$, $E_{\mathrm T}^{\mathrm p}$, and $E_{\mathrm T}^{\Lambda}$ are calculated from
the particle spectra measured by ALICE~\cite{Abelev:2013vea,Abelev:2013xaa} as for the calculation of \ET from the particle spectra. The systematic uncertainties are also propagated assuming that the systematic uncertainties from  different charges of the same particle species (e.g., $\pi^+$ and $\pi^-$) are 100\% correlated and from different species (e.g., $\pi^+$ and K$^+$) are uncorrelated.  The contribution from the $\Sigma^+$, $\Sigma^-$, and $\Sigma^0$ and their antiparticles is determined from the measured $\Lambda$ spectra.  The total contribution from $\Sigma$ species and their antiparticles should be approximately equal to that of the \lam and \alam, but since there are three isospin states for the $\Sigma$, each species carries roughly 1/3 of the \ET that the \lam carries.  Since the $\Sigma^0$ decays dominantly through a \lam and has a short lifetime, the measured \lam spectra include \lam from the $\Sigma^0$ decay.  The ratio of $F$~=~$(E_{\mathrm T}^{\Sigma^{+}}+E_{\mathrm T}^{\Sigma^{-}})/E_{\mathrm T}^{\Lambda}$ is therefore expected to be 0.5.  HIJING~\cite{Wang:1991hta} simulations indicate that $F$~=~0.67 and if the \ET scales with the yield, THERMUS~\cite{Wheaton:2004qb} indicates that $F$~=~0.532.  We therefore use $F$~=~0.585~$\pm$~0.085.

The contribution $E_{\mathrm T}^{\omega,\eta,{\mathrm e}^{\pm},\gamma}$ is calculated using transverse mass scaling for the $\eta$ meson and PYTHIA simulations for the $\omega$, $e^{\pm}$, and $\gamma$, as described earlier. Because most of the \ET is carried by \pikpn, whose contributions appear in both the numerator and the denominator, \ftotal and \fneutral can be determined to high precision, and the uncertainty in \ftotal and \fneutral is driven by $E_{\mathrm T}^{\Lambda}$ and $ E_{\mathrm T}^{\omega,\eta,e^{\pm},\gamma}$.  It is worth considering two special cases.  If all \ET were carried by pions, as is the case at low energy where almost exclusively pions are produced, \Eref{Eq:ftotal2} would simplify to \ftotal = 2/3.  If all \ET were only carried by kaons, (anti)protons, and (anti)neutrons, \Eref{Eq:ftotal2} would simplify to \ftotal = 1/2.

In order to calculate the contribution from the $\eta$ meson and its uncertainties, we assume that the shapes of its spectra for all centrality bins as a function of transverse mass are the same as the pion spectra, using the transverse mass scaling~\cite{Adare:2010fe}, and that the $\eta/\pi$ ratio is independent of the collision system, as observed by PHENIX~\cite{Adare:2010dc}.
We also consider a scenario where the $\eta$ spectrum is assumed to have the same shape as the kaon spectrum, as would be expected if the shape of the $\eta$ spectrum was determined by hydrodynamical flow.   In this case we use the ALICE measurements of $\eta/\pi$ in \pp collisions~\cite{Abelev:2012cn} to determine the relative yields.  We use the  $\eta/\pi$ ratio at the lowest momentum point available, \pT~=~0.5 \GeV, because the \ET measurement is dominated by low momentum particles.  Because no $\omega$ measurement exists, PYTHIA~\cite{Sjostrand:2006za} simulations of \pp collisions were used to determine the relative contribution from the $\omega$ and from all other particles which interact electromagnetically (mainly $\gamma$ and e$^\pm$).  These contributions were approximately 2\% and 1\% of $E_{\mathrm T}^{\pi}$, respectively.  With these assumptions, $E_{\mathrm T}^{\omega,\eta,e^{\pm},\gamma}/E_{\mathrm T}^{\pi}$ = 0.17 $\pm$ 0.11.   The systematic uncertainty on this fraction is dominated by the uncertainty in the $\eta/\pi$ ratio.  We propagate the uncertainties assuming that the \ET from the same particle species are 100\% correlated and that the uncertainties from different particle species are uncorrelated.  

\begin{figure}
\begin{center}
\includegraphics[width=\figurewidth]{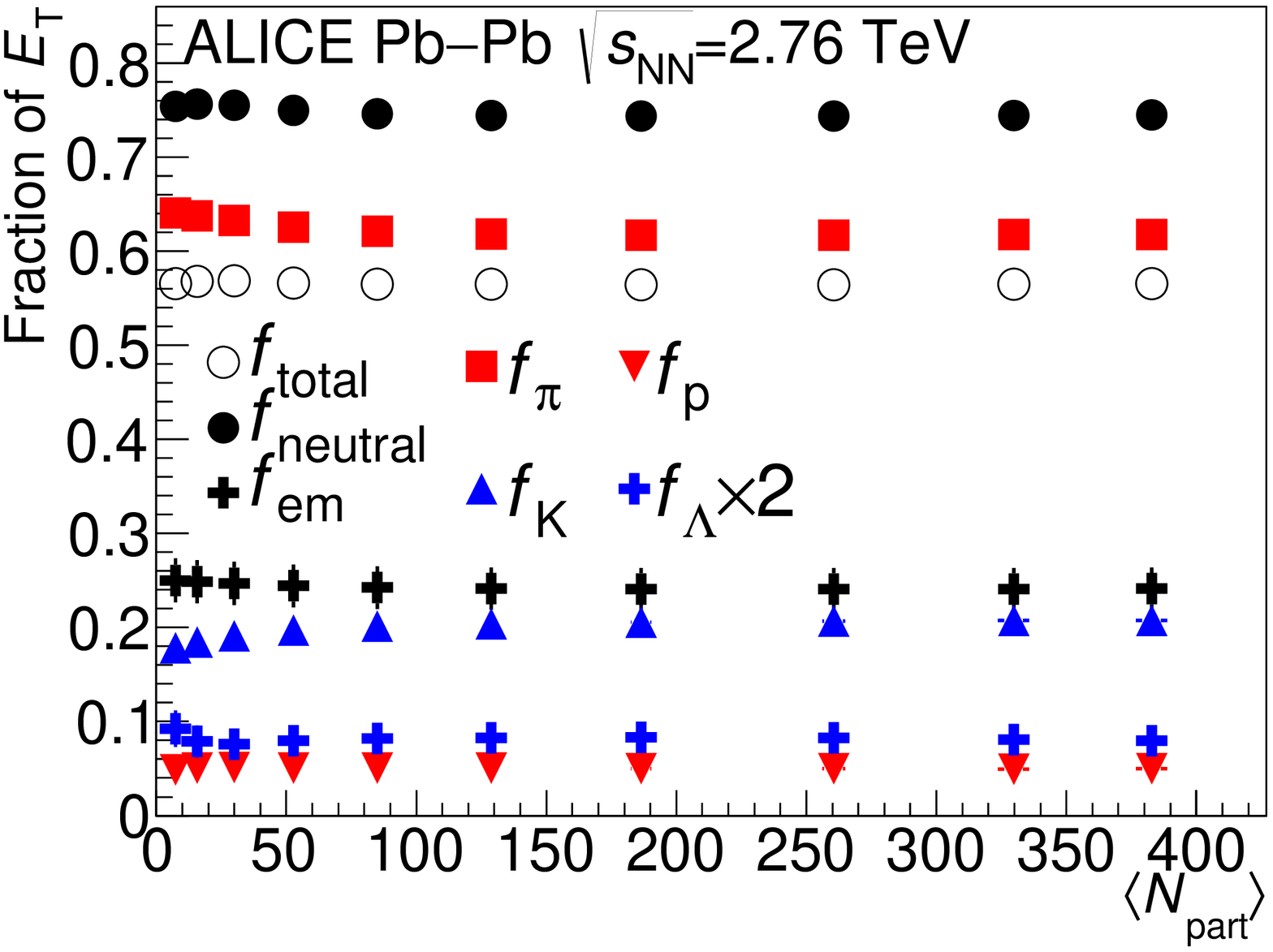}
\caption{Fraction of the total \ET in pions ($f_{\pi}$), kaons ($f_{\mathrm K}$), p and $\overline{p}$ ($f_{{\mathrm p}}$), and $\Lambda$ ($f_{\Lambda}$) and the correction factors \ftotal, \fneutral, and \fem as a function of \Np. The fraction $f_{\Lambda}$ is scaled by a factor of two so that the data do not overlap with those from protons.  Note that \fneutral is the fraction of \EThad measured in the tracking detectors while \ftotal and \fem are the fractions of the total \ET measured in the tracking detectors and the calorimeters, respectively.  The vertical error bars give the uncertainty on the fraction of \ET from the particle yields.}\label{Fig:ETfractions}
\end{center}
\end{figure}

The \fneutral, \ftotal, and \fem = $1-$\ftotal/\fneutral
are shown in \Fref{Fig:ETfractions} along with the fractions of \ET carried by all pions \fpi, all kaons \fK, protons and antiprotons \fp, and $\Lambda$ baryons \flambda versus  \npart.  While there is a slight dependence of the central value on \npart, this variation is less than the systematic uncertainty.  Since there is little centrality dependence, we use \fem = 0.240 $\pm$ 0.027, \fneutral = 0.728 $\pm$  0.017, and \ftotalpaperone, which encompass the entire range for all centralities.  The systematic uncertainty is largely driven by the contribution from $\Lambda$, \otheremparticles since these particles only appear in the denominator of \Erefs{Eq:fneutral2}{Eq:ftotal2}.  The systematic uncertainty on \ftotal is smaller than that on \fneutral because \fneutral only has $E_{\mathrm T}^{\Lambda}$ in the denominator.

These results are independently interesting.  There is little change in the fraction of energy carried by different species with centrality and the changes are included in the \ftotal used for the measurement of \ET.  Additionally, only about 1/4 of the energy is in \ETem, much less than the roughly 1/3 of energy in \ETem at lower energies where most particles produced are pions with the \pizero carrying approximately 1/3 of the energy in the collision.  Furthermore, only about 3.5\% of the \ET is carried by \otheremparticles.  Since charged and neutral pions have comparable spectra, this means that the tracking detectors are highly effective for measuring the transverse energy distribution in nuclear collisions.

\makeatletter{}\subsubsection{\EThadbold distributions}

\begin{table}
\begin{center}
\label{Tab:hadCorrectionSysErrors}
\begin{tabular}{r|c c}
Correction & Value  & \% Rel. uncertainty \\ \hline
\fptcut & 0.9710 $\pm$ 0.0058 & 0.6 \% \\
\fneutral & 0.728 $\pm$  0.017 & 2.3 \% \\
\ftotal & 0.553 $\pm$ 0.010 & 3.0 \% \\%0.559243 ± 0.016564 
\fnotid & 0.982 $\pm$ 0.002 & 0.2 \% \\
\fbkgd & 1.8\% & 0.8\% \\
\eff & 50\% & 5\% \\
\end{tabular}
\end{center}
\caption{Summary of corrections and systematic uncertainties for \EThad and \ET from tracking detectors.  For centrality and \pT independent corrections the correction is listed.  For centrality and \pT dependent corrections, the approximate percentage of the correction is listed.  In addition, the anchor point uncertainty in the Glauber calculations leads to an uncertainty of 0--4\%, increasing with centrality.}\label{Tab:hadCorrectionSysErrors}
\end{table}

\begin{figure}[t]
\begin{center}
\includegraphics[width=\figurewidth]{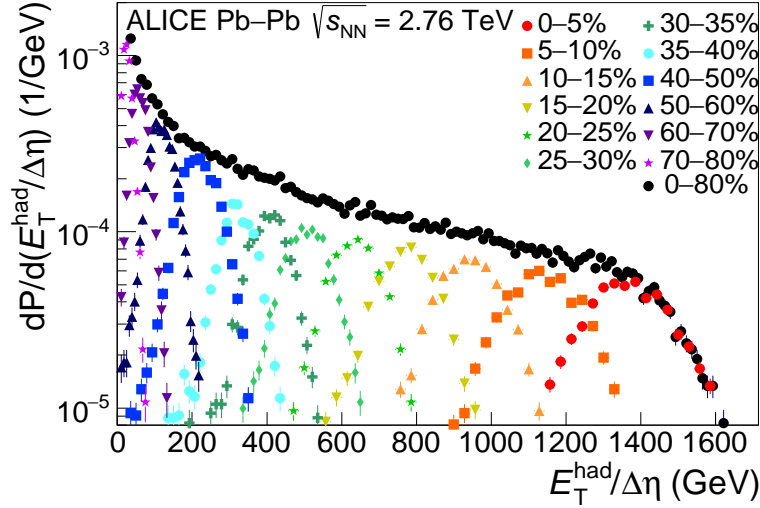}
\caption{Distribution of \EThad measured from \pikp tracks at midrapidity for several centrality classes.  Not corrected for resolution effects.  Only statistical error bars are shown.}
\label{Fig:ETHadDistributions}
\end{center}
\end{figure}

\Figure{Fig:ETHadDistributions} shows the distributions of the reconstructed \EThad measured from \pikp tracks using the method described above for several centralities.   No correction was done for the resolution leaving these distributions dominated by resolution effects.  The mean \EThad is determined from the average of the distribution of \EThad in each centrality class.  

\begin{figure}[th]
\begin{center}
\includegraphics[width=\figurewidth]{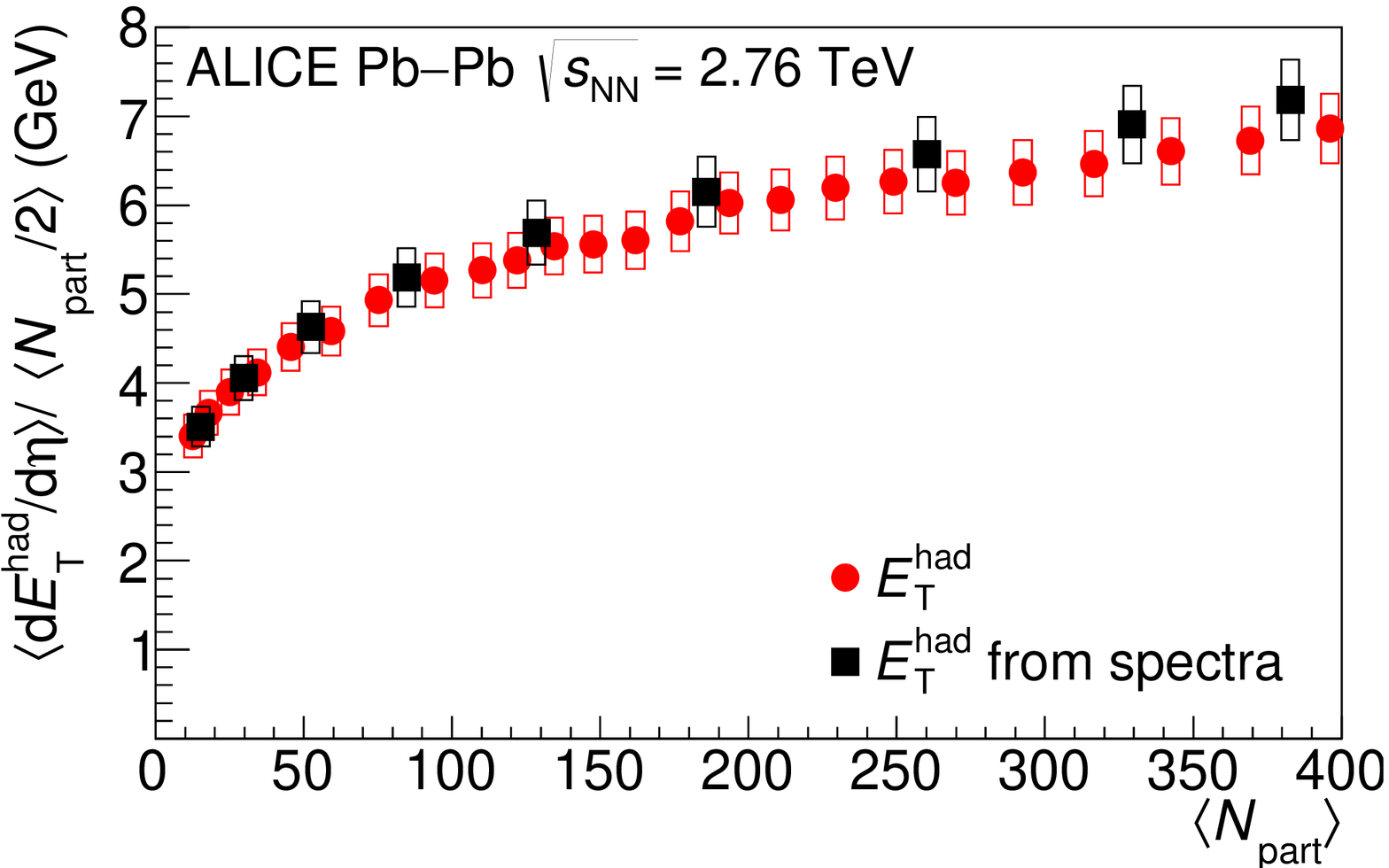}
\caption{Comparison of \meandEthaddetaPerNp versus \npart from the measured particle spectra and as calculated from the tracking detectors.  The boxes indicate the systematic uncertainties.}\label{Fig:ETHadVsNpart}
\end{center}
\end{figure}

\subsection{Calculation of \ETbold and \EThadbold from measured spectra}
We use the transverse momentum distributions (spectra) measured by ALICE~\cite{Abelev:2013vea,Abelev:2013xaa} to calculate \ET and \EThad as a cross check.
We assume that all charge signs and isospin states of each particle carry the same \ET, e.g. $E_{\mathrm T}^{\pi^+}  = E_{\mathrm T}^{\pi^-}  = E_{\mathrm T}^{\pi^0}$, and that the \ET carried by (anti)neutrons equals the \ET carried by (anti)protons.  These assumptions are consistent with the data at high energies where positively and negatively charged hadrons are produced at similar rates and the anti-baryon to baryon ratio is close to one~\cite{PhysRevLett.105.072002,Abbas:2013rua}.  Since the $\Lambda $ spectra~\cite{Abelev:2013xaa} are only measured for five centrality bins, the $\Lambda$ contribution is interpolated from the neighboring centrality bins.  The same assumptions about the contributions of the $\eta$, $\omega$, $\gamma$, and $e^{\pm}$ described in the section on \ftotal and \fneutral are used for these calculations.  The dominant systematic uncertainty on these measurements is due to the single track reconstruction efficiency and is correlated point to point.  The systematic uncertainty on these calculations is not correlated with the calculations of \ET using the tracking detectors because these measurements are from data collected in different years.  The mean \EThad  per \npartpair obtained from the tracking results of \Fref{Fig:ETHadDistributions} are shown as a function of \npart in \Fref{Fig:ETHadVsNpart}, where they are compared with results calculated using the particle spectra measured by ALICE. The two methods give consistent results.  Data are plotted in 2.5\% wide bins in centrality for 0--40\% central collisions, where the uncertainty on the centrality is $<$1\%~\cite{Abelev:2013qoq}.  Data for 40--80\% central collisions are plotted in 5\% wide bins.

\makeatletter{}
\subsection{Electromagnetic calorimeter measurements of \ETembold}\label{Sec:HadEt}

The \ETem is defined as the transverse energy of the particles of category B discussed above, which are the particles measured well by an electromagnetic calorimeter.
While the definition of \ETem includes \allemparticles, the majority of the \ET comes from \pizero $\rightarrow$ $\gamma\gamma$ (85\%) and $\eta$ $\rightarrow$ $\gamma\gamma$ (12\%) decays, meaning that the vast majority of \ETem arises from photons reaching the active area of the electromagnetic calorimeters.  Reconstructed clusters are used for the analysis, with most clusters arising from a single $\gamma$.  Clusters reduce contributions from detector noise to a negligible level, as compared to using tower energies as done by STAR~\cite{Adams:2004cb}.  However, clusters also require additional corrections for the reconstruction efficiency, nonlinearity, and minimum energy reconstructed.  In addition, both the EMCal and the PHOS have limited nominal acceptances so an acceptance correction must be applied.  Backgrounds come from charged hadrons in category A (\pikp), kaon decays into \pizero from both category A (K$^\pm$) and category C (\kzeroshort), neutrons from category D, and particles produced by secondary interactions with the detector material.

The corrected \ETem is given by
\begin{equation}
\frac{{\mathrm d}E_{\mathrm T}^{\mathrm{em}}}{{\mathrm d}\eta} = \frac{1}{\Delta\eta}\frac{1}{f_{\mathrm {acc}}}\frac{1}{f_{\mathrm E_{\mathrm {Tmin}}}}\bigg(\sum\limits_{\mathrm j} \delta_{\mathrm m}\frac{ \sin\theta_{\mathrm j}}{\epsilon_{\gamma} f_{E_{\mathrm {NL}}}} E_{\mathrm j} ~-~E_{\mathrm T}^{\mathrm {bkgd}}\bigg)\label{Eq:EmEt}
\end{equation}
\noindent where j runs over the reconstructed clusters in the calorimeter and $\Delta\eta$ is the pseudorapidity range used in the analysis.  The correction factor \facc corrects for the finite nominal azimuthal detector acceptance, \femin is a correction for the minimum cluster energy used in the analysis, \deltamatched is zero when a cluster is matched to a track and one otherwise, \fereceff is the product of the active acceptance and the reconstruction efficiency in the nominal acceptance of the detector, \fenl is the correction for the nonlinear response of the calorimeter, and \fetbkgd is the sum of the contributions from charged hadrons, kaons, neutrons, and particles created by secondary interactions. 
These correction factors are discussed below and their systematic uncertainties are summarized in \Tref{Tab:emEtCorrectionSysErrors}.  All of the systematic uncertainties except for that due to the background subtraction are correlated point to point.  Systematic uncertainties on measurements of \ETem from the EMCal and the PHOS and calculations of \ETem from the spectra are not correlated.  Systematic uncertainties on hybrid measurements are dominated by systematic uncertainties on \EThad and are therefore dominantly correlated point to point and with the tracking detector measurements.

\subsubsection{Acceptance correction \faccbold and cluster reconstruction efficiency \fereceffbold}
The correction for the acceptance is divided into two parts, the correction due to the nominal acceptance of the detector and the correction due to limited acceptance within the nominal acceptance of the detector due to dead regions and edge effects.  
To reduce edge effects, clusters in the PHOS are restricted to $|\eta| <$~0.1 and in the EMCal to $|\eta| <$~0.6.  
The correction \facc accounts for the limited nominal acceptance in azimuth and is therefore 5/18 for the EMCal, which has a nominal acceptance of 100$^\circ$, and 1/6 for the PHOS, which has a nominal acceptance of 60$^\circ$. It does not correct for acceptance effects due to dead regions in the detector or for noisy towers omitted from the analysis. This is accounted
for by the cluster reconstruction efficiency$\times$acceptance within the nominal detector acceptance,  \fereceff, calculated from \sims using photons from the decay of the \pizero\ meson.  The efficiency is calculated as a function of the energy of the cluster.

\subsubsection{Minimum cluster energy \feminbold}
There is a minimum energy for usable clusters analogous to the minimum \pT~in the acceptance of the tracking detectors.
A threshold of 250 MeV for PHOS and 300 MeV for the EMCal is applied.  These energies are above the peak energy for minimum ionizing particles (MIPs), reducing the background correction due to charged hadrons.  We apply the threshold in \ET rather than energy because it simplifies the calculation of the correction for this threshold and its systematic uncertainty.  We use the charged pion spectra to calculate the fraction of \ETem below these thresholds.  PYTHIA is used to simulate the decay kinematics and the measured charged pion spectra are used to determine the fraction of \ET from pions within the acceptance.  As for the calculation of \ftotal for the measurement of \EThad described above, we assume transverse mass scaling to determine the shape of the $\eta$ spectrum and the $\eta/\pi$ ratio measured by ALICE~\cite{Abelev:2012cn} to estimate the contribution of the $\eta$ meson to \femin.  The uncertainty on the shape of the charged pion spectrum and on the $\eta/\pi$ ratio is used to determine the uncertainty on \femin.  
This correction is centrality dependent and ranges from 0.735 to 0.740 for the PHOS and from 0.640 to 0.673 for the EMCal with a systematic uncertainty of 3.5--5\%.

\subsubsection{Nonlinearity correction \fenlbold and energy scale uncertainty}
For an ideal calorimeter the signal observed is proportional to the energy.  In practice, however, there is a slight deviation from linearity in the signal observed, particularly at low energies.  A nonlinearity correction is applied to take this into account.  For the EMCal this deviation from linearity reaches a maximum of about 15\% for the lowest energy clusters used in this analysis.  The systematic uncertainty for the EMCal is determined by comparing the nonlinearity observed in test beam and the nonlinearity predicted by \sims and reaches a maximum of about 5\% for the lowest energy clusters.  The PHOS nonlinearity is determined by comparing the location of the \pizero mass peak to \sims and cross checked using the energy divided by the momentum for identified electrons.  The systematic uncertainty is derived from the accuracy of the location of the \pizero mass peak.  The nominal correction is about 1\% with a maximum systematic uncertainty of around 3\% for the lowest energy clusters.  The raw \ETem is calculated with the maxima and minima of the nonlinearities and the difference from the nominal value is assigned as a systematic uncertainty.  The final systematic uncertainty on the measurement with the EMCal due to nonlinearity is about 0.8\% and 1.3\% for the PHOS.  For both the PHOS and the EMCal, the energy scale uncertainty was determined by comparing the location of the \pizero mass peak and the ratio of energy over momentum for electrons.  This systematic uncertainty is 2\% for the EMCal~\cite{Abelev:2013fn} and 0.5\% for the PHOS~\cite{Abelev:2014ypa}.

\subsubsection{Background \fetbkgdbold}
Charged particles (category A) are the largest source of background in \ETem.  Clusters matched to tracks are omitted from the analysis.  The track matching efficiency determined from \sims is combined with information from clusters matched to tracks to calculate the number and mean energy of remaining deposits from charged particles.  The systematic uncertainty on this contribution comes from the uncertainty on the track matching efficiency and the uncertainty in the mean energy.  The former is dominated by the uncertainty on the single track reconstruction efficiency and the latter is determined by comparing central and peripheral collisions, assuming that the energy of clusters matched to tracks in central collisions may be skewed by overlapping clusters due to the high occupancy.

The background contributions from both charged kaons (category A) through their K$^{\pm}$ $\rightarrow$ X\pizero decays and \kzeroshort (category C) through its \kzeroshort $\rightarrow$ \pizero\pizero decay are non-negligible.  
The amount of energy deposited by a kaon as a function of \pT is determined using \sims.  This is combined with the kaon spectra measured by ALICE~\cite{Abelev:2013vea} to calculate the energy deposited in the calorimeters by kaons.
The systematic uncertainty on the background from kaons is determined by varying the yields within the uncertainties of the spectra.
Contributions from both neutrons and particles from secondary interactions are determined using \sims.  The systematic uncertainty on these contributions is determined by assuming that they scale with either the number of tracks (as a proxy for the number of charged particles) or with the number of calorimeter clusters (as a proxy for the number of neutral particles).

The background contribution is centrality dependent and ranges from 61\% to 73\% with both the background and its systematic uncertainty dominated by contributions from charged hadrons.  This correction is so large because \ETem comprises only about 25\% of the \ET in an event while \pikp carry roughly 57\% of the \ET in an event.

\subsubsection{Acceptance effects}
The limited calorimeter acceptance distorts the distribution of \ETem for events with very low \ETem because it is difficult to measure the mean \ET when the mean number of clusters observed is small (about 1--10).  
While it is possible to correct for acceptance, this was not done since the measurement of \ET from the tracking method has the highest precision.  The hybrid method using both the calorimeters and the tracking detectors is therefore restricted to the most central collisions where distortions of the \ETem distribution are negligible.

\makeatletter{}
\subsubsection{\ETembold distributions}
\begin{table}
\begin{center}
\label{Tab:emEtCorrectionSysErrors}
\begin{tabular}{r|c c | c c}
 &  \multicolumn{2}{c}{PHOS} & \multicolumn{2}{c}{EMCal} \\ 
 & Correction & Uncertainty & Correction & Uncertainty\\ \hline
 \facc & 6 & 0 & 3.6 & 0 \\
Energy scale & -- & 0.5\% & -- & 2\% \\
\fereceff & 40\% & 5\% & 80\% & 5\% \\
\femin & 0.735 -- 0.740 & 3.5\% & 0.64 -- 0.673 & 4.1 -- 5.0\% \\
 \fenl & $<$ 0.5\% & 1.3\% & $<$ 5\% & 0.8\% \\
\fbkgdemet & 0.616 -- 0.753 & 9 -- 20\% & 0.659 -- 0.732 & 8 -- 13\% \\ \hline
\ETem & -- & 10 -- 20\%  & -- & 10 -- 15\% \\
\end{tabular}
\end{center}
\caption{Summary of corrections and systematic uncertainties for \ETem.  The approximate size of the correction is listed for \fereceff and the ranges are listed for centrality dependent corrections.  
The fraction \fbkgdemet = $E_{\mathrm T}^{\mathrm {bkgd}}/E_{\mathrm T}^{\mathrm {raw}}$ where $E_{\mathrm T}^{\mathrm {raw}} = \sum\limits_{\mathrm j} \delta_{\mathrm m}\frac{ \sin\theta_{\mathrm j}}{\epsilon_{\gamma} f_{E_{\mathrm {NL}}}} E_{\mathrm j}$ is given in order to compare \fetbkgd across centralities.  In addition, the anchor point uncertainty in the Glauber calculations leads to an uncertainty of 0--4\%, increasing with centrality.
}\label{Tab:emEtCorrectionSysErrors}
\end{table}

\begin{figure}[t]
\begin{center}
\includegraphics[width=\figurewidth]{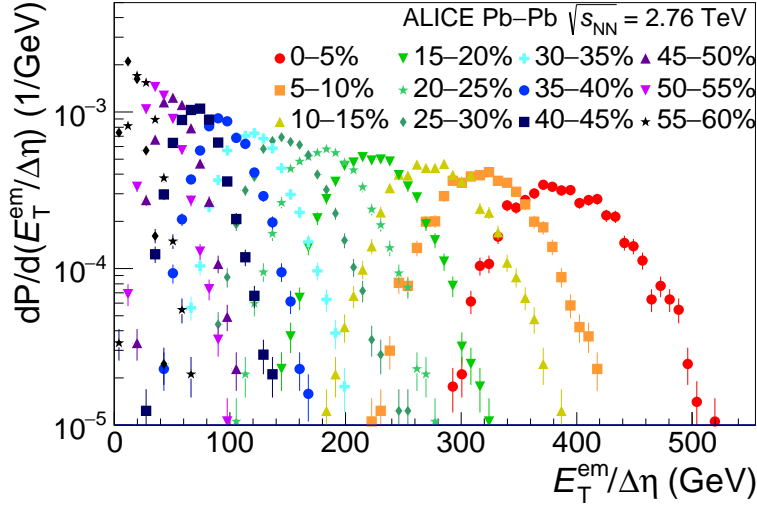}
\caption{Distribution of \ETem  measured with the EMCal at midrapidity for several centrality bins.  Not corrected for resolution effects.  Only statistical error bars are shown.}\label{Fig:ETEmDistributionEmcal}
\end{center}
\end{figure}

\begin{figure}[t]
\begin{center}
\includegraphics[width=\figurewidth]{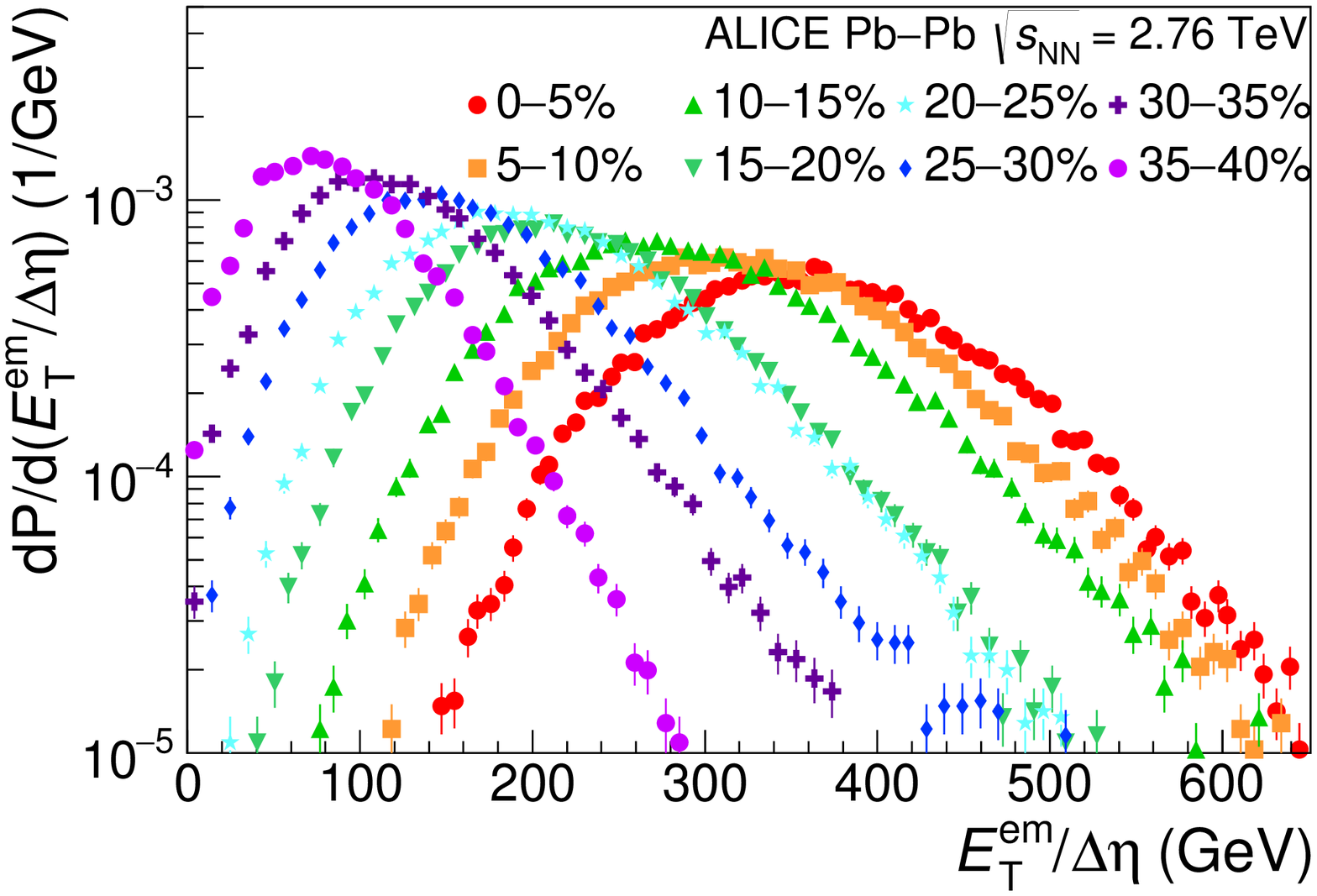}
\caption{Distribution of \ETem  measured with the PHOS at midrapidity for several centrality bins.  Not corrected for resolution effects.  Only statistical error bars are shown.}\label{Fig:ETEmDistributionPhos}
\end{center}
\end{figure}

\begin{figure}[t]
\begin{center}
\includegraphics[width=\figurewidth]{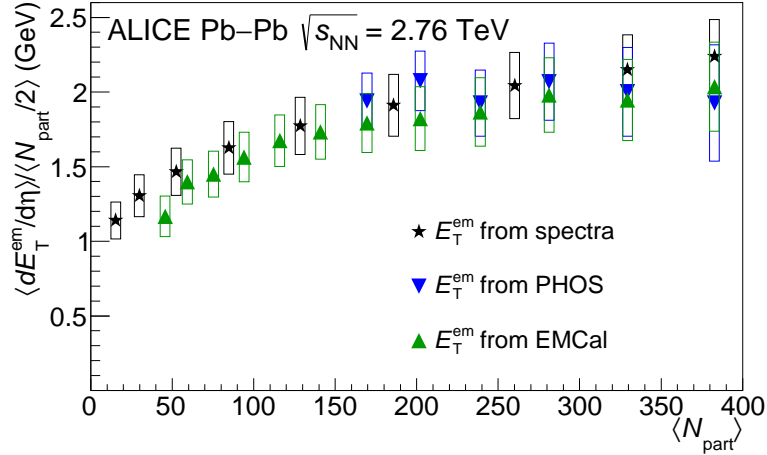}
\caption{Comparison of \meandEtemdetaPerNp versus \npart at midrapidity from the PHOS, from the EMCal, and as calculated from the measured pion spectra.  The boxes indicate the systematic uncertainties.}\label{Fig:ETEmComparisons}
\end{center}
\end{figure}

%\Figure{Fig:ETHadDistributions} shows the distributions of the reconstructed \EThad measured from \pikp tracks using the method described above for several centralities.
\Figure{Fig:ETEmDistributionEmcal} shows the distributions of the reconstructed \ETem measured using the EMCal and \Fref{Fig:ETEmDistributionPhos} shows the distributions of the reconstructed \ETem measured using the PHOS. No resolution correction was applied for the resolution leaving the distributions in \Fref{Fig:ETEmDistributionEmcal} and \Fref{Fig:ETEmDistributionPhos} dominated by resolution effects.  The resolution is primarily determined by the finite acceptance of the detectors in azimuth, limiting the fraction of \ETem sampled by the calorimeter.  The distributions are broader for PHOS than EMCal because of the smaller azimuthal acceptance of the PHOS.  The mean \ETem is determined from the average of the distribution of \ETem in each centrality bin. The \ETem per \Np pair measured using the electromagnetic calorimeters is compared to that calculated using the measured pion spectra in \Fref{Fig:ETEmComparisons}, demonstrating that these methods lead to comparable results.  The \ETem calculated from the spectra is determined using the same ratio of $E_{\mathrm T}^{\omega,\eta,{\mathrm e}^{\pm},\gamma}/E_{\mathrm T}^{\pi}$ = 0.171 $\pm$ 0.110 for all centralities.

\section{Results}\label{Sec:Results}
\makeatletter{}

\begin{figure}
\begin{center}
\includegraphics[width=\figurewidth]{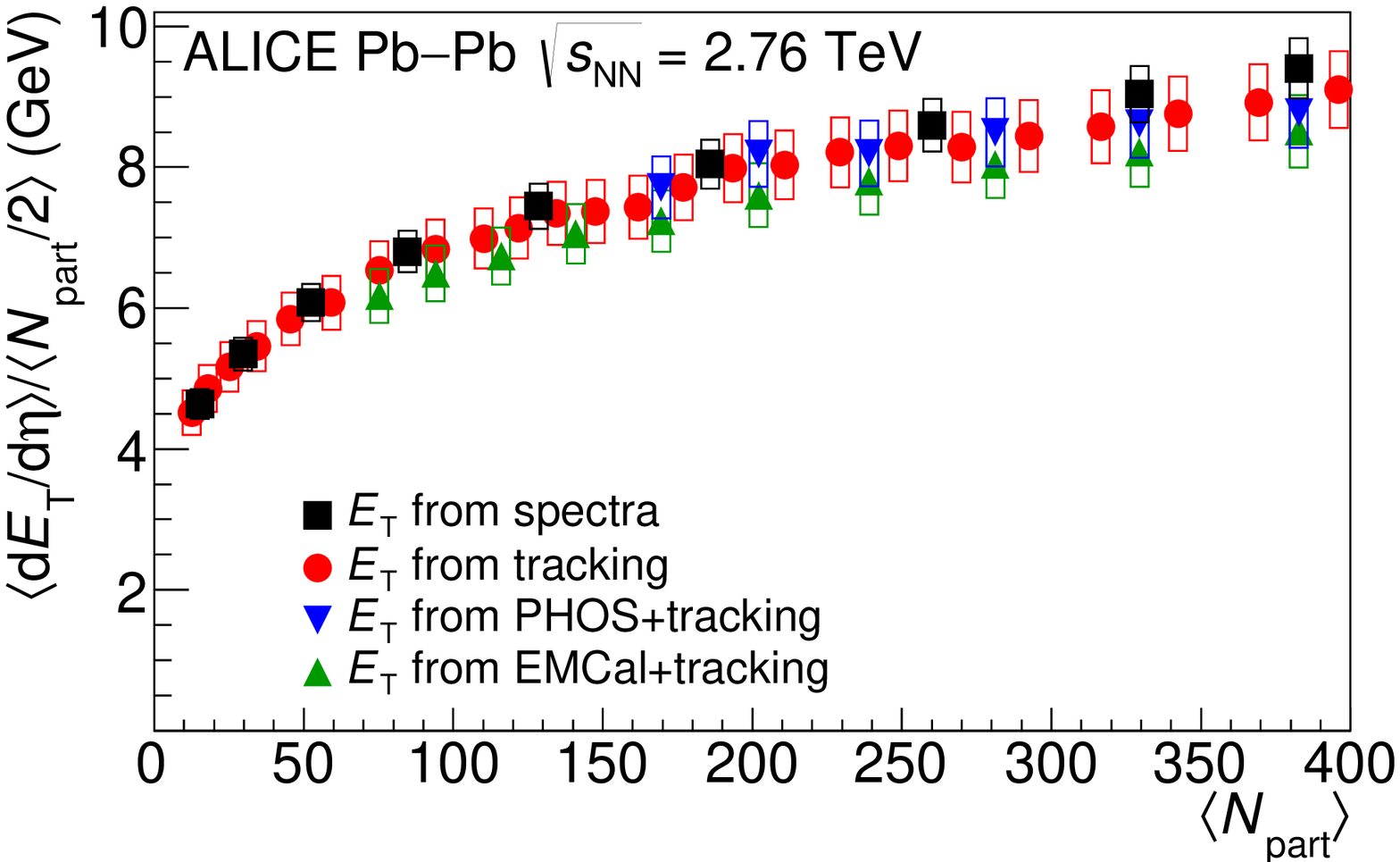}
\caption{Comparison of total \meandEtdetaPerNp versus \npart at midrapidity using tracking detectors, using EMCal+tracking, using PHOS+tracking, and as calculated from the measured particle spectra.  The boxes indicate the systematic uncertainties.}\label{Fig:ETTotVsNpart}
\end{center}
\end{figure}

The \meandEtdetaPerNp versus \npart is shown in \Fref{Fig:ETTotVsNpart} using tracking detectors, using EMCal+tracking, using PHOS+tracking, and as calculated from the measured particle spectra.  All methods lead to comparable results, although the systematic errors are largely correlated due to the dominant correction from the tracking inefficiency.  The determination of \Npart and its uncertainties are calculated using a Glauber model as in~\cite{Abelev:2013qoq,Loizides:2016djv} and the uncertainties on \Npart are added in quadrature to the uncertainties on \ET.

 As discussed above, the small number of clusters observed in the calorimeters in peripheral collisions make acceptance corrections difficult.  Since the measurements with the tracking detectors alone has higher precision, only these measurements are used in the following.
 
\begin{figure}[t]
\begin{center}
\includegraphics[width=\figurewidth]{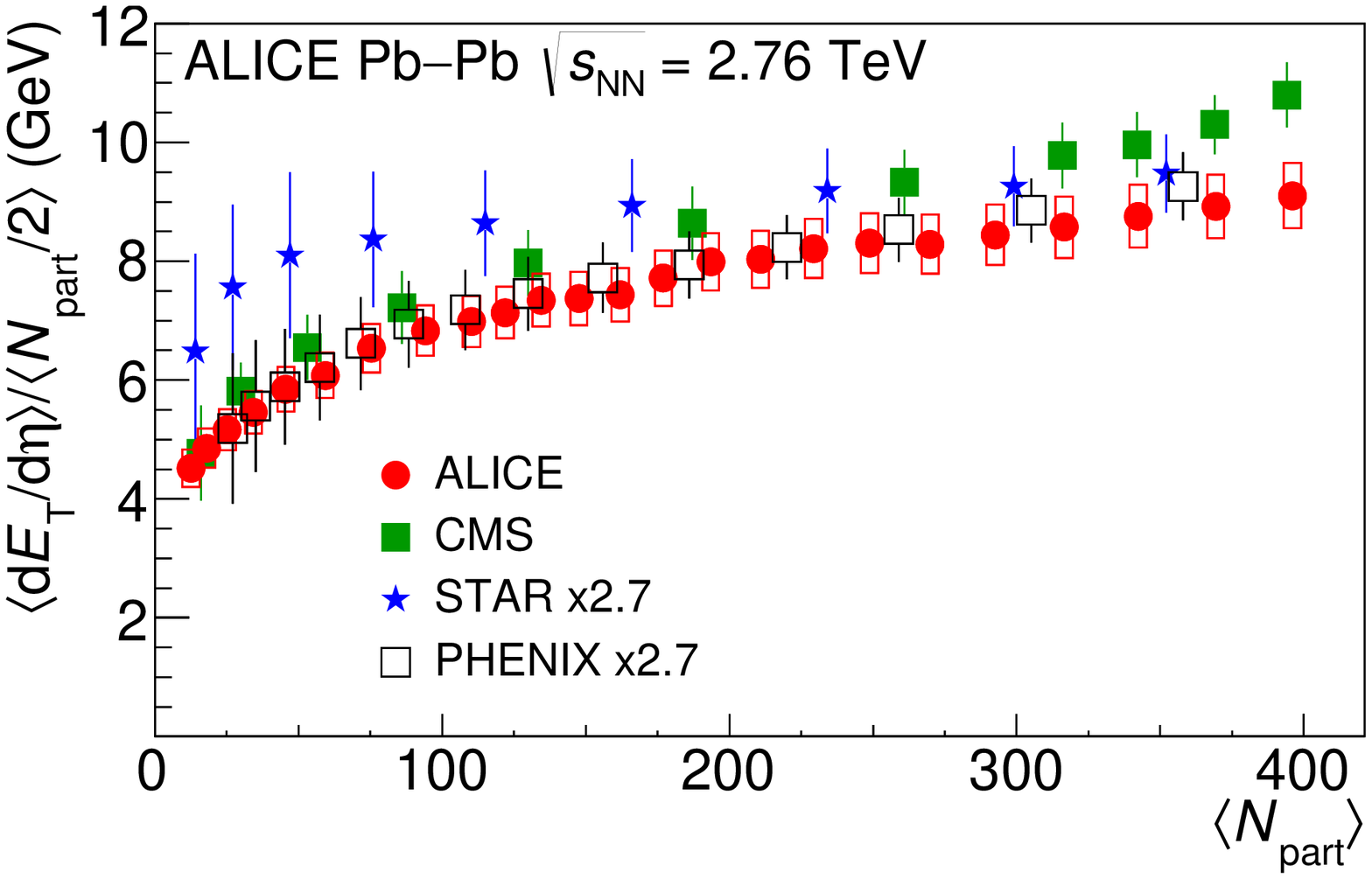}
\caption{Comparison of \meandEtdetaPerNp  at midrapidity in \Pb collisions at \sNN = 2.76 TeV from CMS~\cite{CMSET:2012} and ALICE and in \Au collisions at \sNN = 200 GeV from STAR~\cite{Adams:2004cb} and PHENIX~\cite{Adler:2004zn,Adcox:2001ry}.  Data from RHIC were scaled by a factor of 2.7 for comparison of the shapes. 
The boxes indicate the systematic uncertainties.}
\label{Fig:ETTotVsNpartExpComp}
\end{center}
\end{figure}

\makeatletter{}\Figure{Fig:ETTotVsNpartExpComp} compares \meandEtdetaPerNp versus \npart in \Pb collisions at \sNN = 2.76 TeV from CMS~\cite{CMSET:2012} and ALICE, and in \Au collisions at \sNN = 200 GeV from STAR~\cite{Adams:2004cb} and PHENIX~\cite{Adler:2004zn,Adcox:2001ry}.  
Data from RHIC have been scaled by a factor of 2.7 for comparison of the shapes. The factor of 2.7 is approximately the ratio of \meanpTmeandNchdeta  at the LHC~\cite{Abelev:2013vea} to that at RHIC~\cite{Abelev:2008ab,Adler:2003cb}.  
The shapes observed by ALICE and PHENIX are comparable for all \npart.  STAR measurements are consistent with PHENIX measurements for the most central collisions and above the PHENIX measurements, although consistent within systematic uncertainties, for more peripheral collisions.  
CMS measurements are consistent with ALICE measurements for peripheral collisions but deviate beyond the systematic uncertainties for more central collisions.  The \ET in 0--5\% \Pb collisions is \centralet and the \ET per participant is \centraletnpart, two standard deviations below the value observed by CMS~\cite{CMSET:2012}.  All methods resulted in a lower \ET than that reported by CMS, although the systematic errors on the measurements are significantly correlated.  One possible explanation of
the differences is that the corrections for the CMS calorimetry measurement are determined by Monte Carlo~\cite{CMSET:2012} while the corrections for the ALICE measurement are mainly data-driven.

\begin{figure}[t]
\begin{center}
\includegraphics[width=\figurewidth]{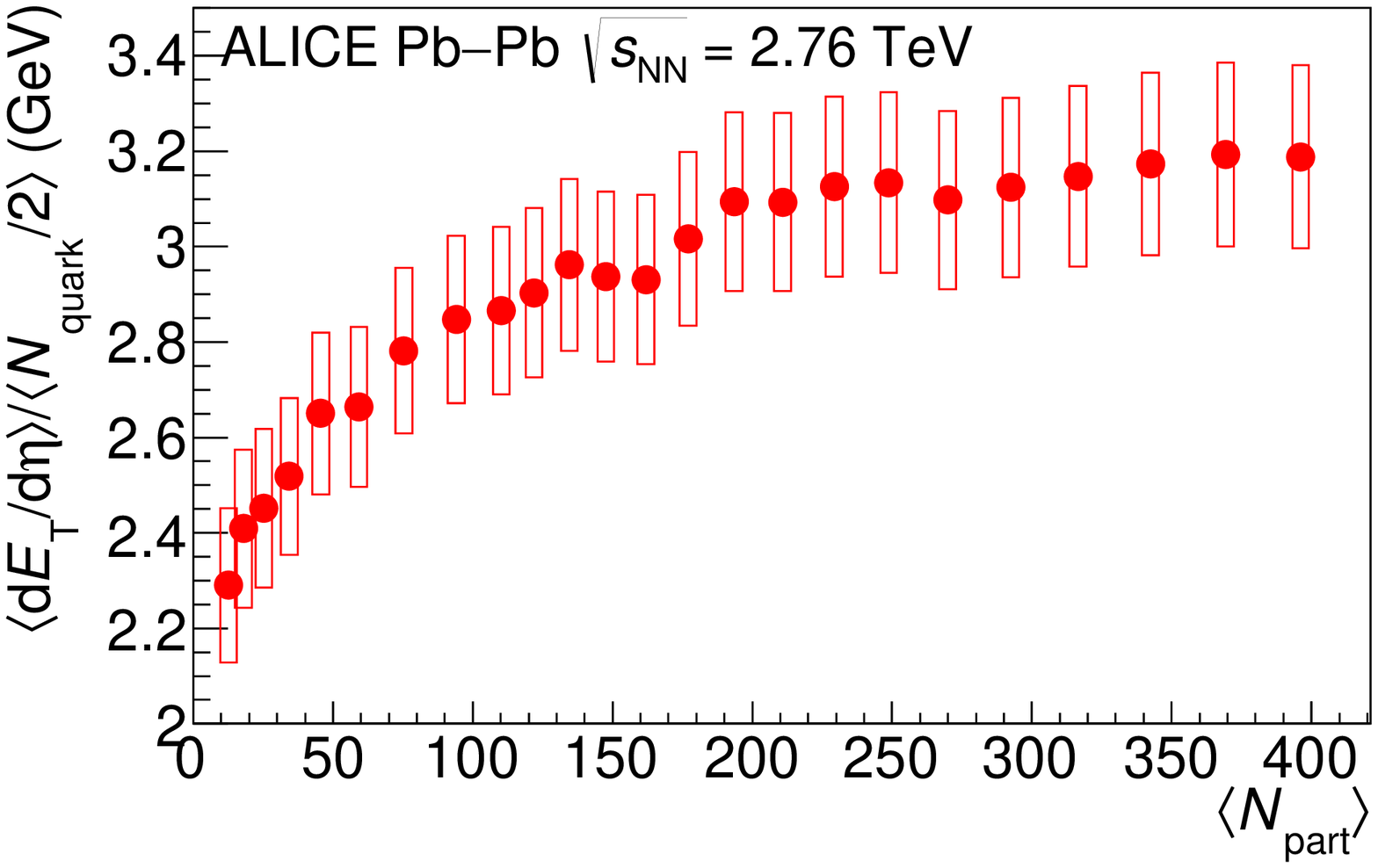}
\caption{Measurements of \meandEtdetaPerNq versus \npart at midrapidity in \Pb collisions at \sNN = 2.76 TeV.  Note the suppressed zero.  The boxes indicate the systematic uncertainties.}\label{Fig:ETTotVsNquarkExpComp}
\end{center}
\end{figure}

PHENIX~\cite{Adler:2013aqf} reported 
that while \meandEtdeta scaled by \Np has a pronounced centrality dependence, as seen in \Fref{Fig:ETTotVsNpartExpComp}, \meandEtdeta scaled by the number of constituent quarks, \Nq, \meandEtdetaPerNq shows little centrality dependence within the systematic uncertainties for collisions at \sNN = 62.4 -- 200 GeV.  This indicates that \ET might scale linearly with the number of quarks participating in the collision rather than the number of participating nucleons. 
\Figure{Fig:ETTotVsNquarkExpComp} shows \meandEtdetaPerNq as a function of \npart. To calculate \Nq the standard Monte Carlo Glauber technique~\cite{Miller:2007} has been used with the following Woods-Saxon nuclear density parameters: radius parameter $R_{\mathrm {WS}} = 6.62 \pm 0.06$ fm, diffuseness $a = 0.546 \pm 0.010$ fm, and hard core $d_{\mathrm {min}} = 0.4 \pm 0.4$ fm. The three constituent quarks in each nucleon have been sampled from the nucleon density distribution $\rho_{\mathrm {nucleon}} = \rho_0 e^{-ar}$ with $a = 4.28 fm^{-1}$ using the method developed by PHENIX~\cite{Adare:2015bua}. The inelastic quark-quark cross section at \sNN = 2.76 TeV was found to be $\sigma_{\mathrm{qq}}^{\mathrm{inel}} = 15.5 \pm 2.0$ mb corresponding to $\sigma_{\mathrm{NN}}^{\mathrm{inel}} = 64 \pm 5$ mb~\cite{Abelev:2013qoq}.   The systematic uncertainties on the \nquark calculations were determined following the procedure described in~\cite{Abelev:2013qoq,Loizides:2016djv}. Unlike at RHIC, we observe an increase in \meandEtdetaPerNq with centrality below \npart~$\approx$200.

\begin{figure}[t]
\begin{center}
\includegraphics[width=\figurewidth]{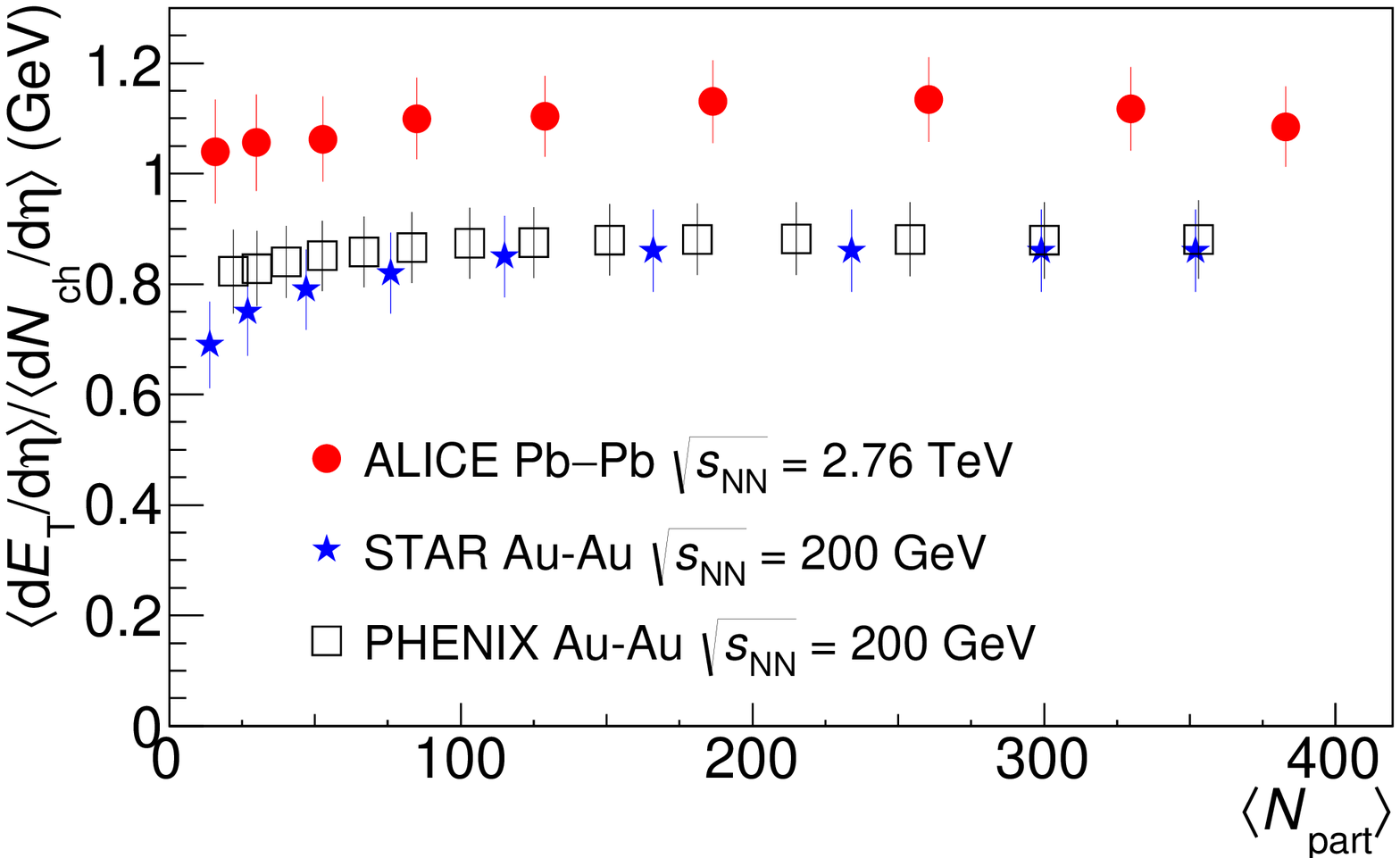}
\caption{Comparison of \meandEtdetaOverdNchdeta versus \npart at midrapidity in \Pb collisions at \sNN~=~2.76~TeV from ALICE and in \Au collisions at \sNN = 200 GeV from STAR~\cite{Adams:2004cb} and PHENIX~\cite{Adler:2004zn,Adcox:2001ry}.}\label{Fig:ETOverNChVsNpartExpComp}
\end{center}
\end{figure}

\Figure{Fig:ETOverNChVsNpartExpComp} shows \meandEtdetaOverdNchdeta, a measure of the average transverse energy per particle, versus \npart in \Pb collisions at \sNN = 2.76 TeV from ALICE, and in \Au collisions at \sNN~=~200~GeV from STAR~\cite{Adams:2004cb} and PHENIX~\cite{Adler:2004zn,Adcox:2001ry}.  No centrality dependence is observed within uncertainties at either RHIC or LHC energies.  The \meandEtdetaOverdNchdeta increases by a factor of approximately 1.25 from \sNN = 200 GeV~\cite{Adler:2004zn,Adcox:2001ry,Adams:2004cb} to \sNN = 2.76 TeV.
% , comparable to the increase in \meanpT from \sNN = 200 GeV~\cite{Abelev:2008ab,Adler:2003cb} to \sNN = 2.76 TeV~\cite{Abelev:2013bla}.  
% The average transverse momentum, \meanpT, also shows little dependence on the charged-particle multiplicity except for peripheral collisions~\cite{Abelev:2013bla}. 
% The absence of a strong centrality dependence in \meandEtdetaOverdNchdeta is consistent with the development of radial flow seen in the spectra of identified particles~\cite{Abelev:2013vea} assuming kinetic energy is conserved during the hydrodynamic expansion.
% While the increasing radial flow with increasing charged particle multiplicity~\cite{Abelev:2013vea} can lead to a higher \meanpT~\cite{Abelev:2013bla,Abelev:2013vea}, kinetic energy should be conserved during the hydrodynamic expansion.  \Figure{Fig:ETOverNChVsNpartExpComp} is consistent with those expectations.
%Kris's suggestion
%   "... 1.25 from \sqrt{s_{NN}}=200 GeV to \sqrt{s_{NN}}=2.76 TeV. 
This is comparable to the increase in \meanpT from \sNN = 200 GeV to \sNN = 2.76 TeV, which also shows little dependence on the charged-particle multiplicity except in peripheral collisions~\cite{Abelev:2013bla}. The absence of a strong centrality dependence is consistent with the development of radial flow seen in the spectra of identified particles~\cite{Abelev:2013bla} where the kinetic energy is conserved during the hydrodynamic expansion instead of producing a higher \meanpT.

\begin{figure}[t]
\begin{center}
\includegraphics[width=\figurewidth]{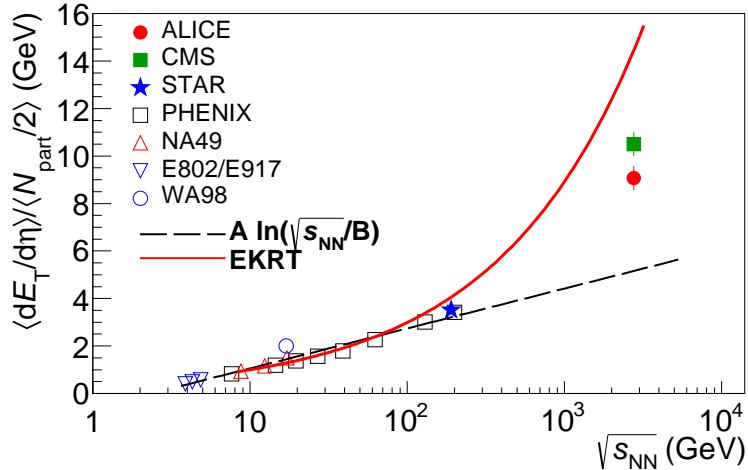}
\caption{Comparison of \meandEtdetaPerNp at midrapidity versus \sNN in 0--5\% central \Pb collisions at \sNN = 2.76 TeV from ALICE and CMS~\cite{CMSET:2012} and central collisions at other energies~\cite{Adams:2004cb,Adler:2004zn,Adare:2015bua} at midrapidity.  All measurements are from 0--5\% central collisions except the NA49 data, which are from 0--7\% collisions.}\label{Fig:ETVsSqrtS}
\end{center}
\end{figure}

\Figure{Fig:ETVsSqrtS} shows a comparison of \meandEtdetaPerNp versus \sNN in 0--5\% central \Pb collisions at \sNN = 2.76 TeV from ALICE and CMS~\cite{CMSET:2012} and central collisions at other energies~\cite{Adams:2004cb,Adler:2004zn,Adare:2015bua} at midrapidity.   The data are compared to an extrapolation from lower energy data~\cite{Adler:2004zn}, which substantially underestimates the \meandEtdetaPerNp at the LHC.  The data are also compared to the EKRT model~\cite{Eskola:1999fc,Renk:2011gj}.  The EKRT model combines perturbative QCD minijet production with gluon saturation and hydrodynamics.  The EKRT calculation qualitatively describes the \sNN dependence at RHIC and SPS energies~\cite{Adams:2004cb}.  However, at LHC energies EKRT overestimates \ET substantially, indicating that it is unable to describe the collision energy dependence.

\begin{figure}
\begin{center}
\rotatebox{0}{\resizebox{\figurewidth}{!}{
        \includegraphics{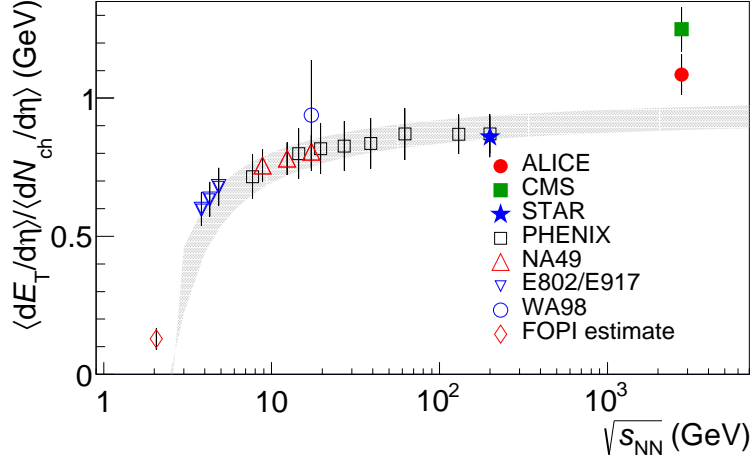}
}}\caption{Comparison of \meandEtdetaOverdNchdeta  at midrapidity versus \sNN in \Pb collisions at \sNN = 2.76 TeV from ALICE and measurements at other energies~\cite{E802Et:1992,Ahle:1994nk,Margetis:1994tt,Aggarwal:2000bc,Adler:2004zn,Adams:2004cb,Adare:2015bua}.  The band shows the extrapolation from lower energies with the width representing the uncertainty on the fit~\cite{Adler:2004zn}.}\label{Fig:ETOverNchVsSqrtS}
\end{center}
\end{figure}

\Figure{Fig:ETOverNchVsSqrtS} shows a comparison of \meandEtdetaOverdNchdeta versus \sNN in 0-5\% central \Pb collisions at \sNN = 2.76 TeV from ALICE and in central collisions at other energies.  Previous measurements indicated that \meandEtdetaOverdNchdeta had either saturated at RHIC energies or showed only a weak dependence on \sNN~\cite{Adler:2004zn,Adams:2004cb,Adare:2015bua}.  An empirical extrapolation of the data to LHC energies assuming that both \ET and \Nch have a linear dependence on \sNN predicted that \meandEtdetaOverdNchdeta would be 0.92 $\pm$ 0.06~\cite{Adler:2004zn} and we observe 1.06 $\pm$ 0.05.  
Increasing the incident energy increases both the particle production and the mean energy per particle at LHC energies, in contrast to lower energies (\sNN = 19.6 -- 200 GeV) where increasing the incident energy only led to increased particle production~\cite{Adler:2004zn}.

\begin{figure}[t]
\begin{center}
\includegraphics[width=\figurewidth]{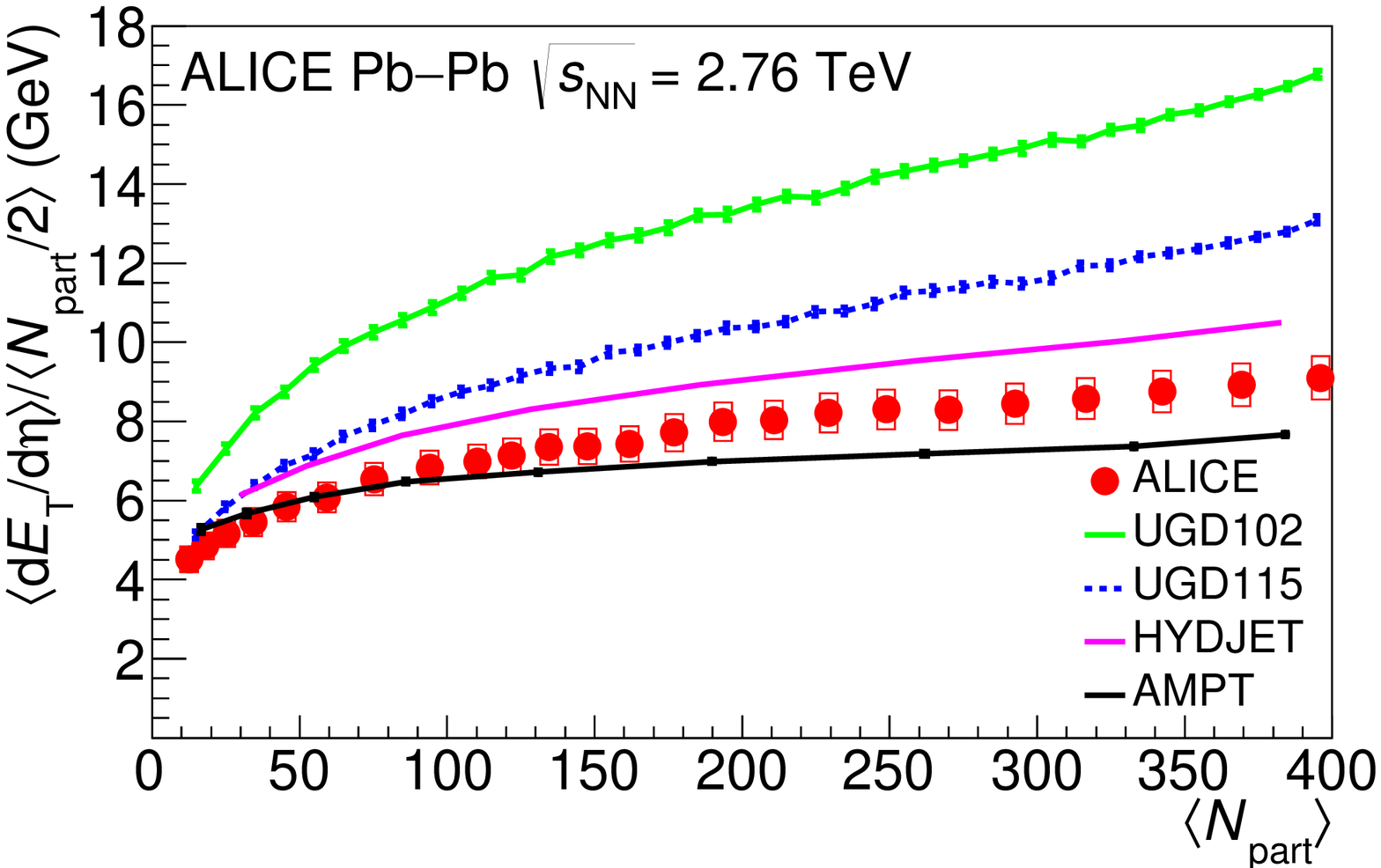}
\caption{Comparison of \meandEtdetaPerNp versus \npart  at midrapidity to AMPT~\cite{Lin:2000cx}, HYDJET 1.8~\cite{Lokhtin:2005px}, and UDG~\cite{Albacete:2012xq}. The boxes indicate the systematic uncertainties.}\label{Fig:ETTotVsNpartWithModels}
\end{center}
\end{figure}

\Figure{Fig:ETTotVsNpartWithModels} shows a comparison of \meandEtdetaPerNp versus \npart to various models.  AMPT~\cite{Lin:2000cx} is a Monte Carlo event generator which builds on HIJING~\cite{Wang:1991hta}, adding explicit interactions between initial minijet partons and final state hadronic interactions.  HYDJET 1.8~\cite{Lokhtin:2005px} is a Monte Carlo event generator that introduces jet quenching via gluon bremsstrahlung to PYTHIA~\cite{Sjostrand:2006za} events.  The HYDJET values use the \ET from HYDJET and the \npart from ALICE, similar to comparisons to HYDJET by CMS~\cite{CMSET:2012}.  The curves labeled UDG are calculations from a Color Glass Condensate model~\cite{Albacete:2012xq} with different normalization K factors.  None of the available models is able to describe the data very well, but we find that AMPT does best in describing the shape and level of \meandEtdetaPerNp.  HYDJET describes the relative shape changes as a function of centrality as well as AMPT, but overestimates \meandEtdetaPerNp.  Both CGC calculations overestimate \meandEtdetaPerNp and predict a larger increase as a function of centrality than is observed in the data.

The volume-averaged energy density $\epsilon$ can be estimated from \meandEtdeta using the following expression~\cite{Bjorken:1982qr}
\begin{equation}
 \epsilon = \frac{1}{\mathrm A c \tau_0} {\mathrm J} \Big\langle \frac{{\mathrm d}E_{\mathrm T}}{{\mathrm d}\eta} \Big\rangle \label{Eqtn:Bjorken}
\end{equation}

\noindent where A is the effective transverse collision area, $c$ is the speed of light, J is the Jacobian for the transformation between \meandEtdeta and \meandEtdy, and $\tau_0$ is the formation time.   The Jacobian is calculated from the measured particle spectra~\cite{Abelev:2013vea,Abelev:2013xaa}.  While J has a slight centrality dependence, it is smaller than the systematic uncertainty so a constant Jacobian of J = 1.12 $\pm$ 0.06 is used.  The formation time of the system $\tau_0$ is highly model dependent and we therefore report $\epsilon\tau_{0}$.

\begin{figure}[t]
\begin{center}
\includegraphics[width=\figurewidth]{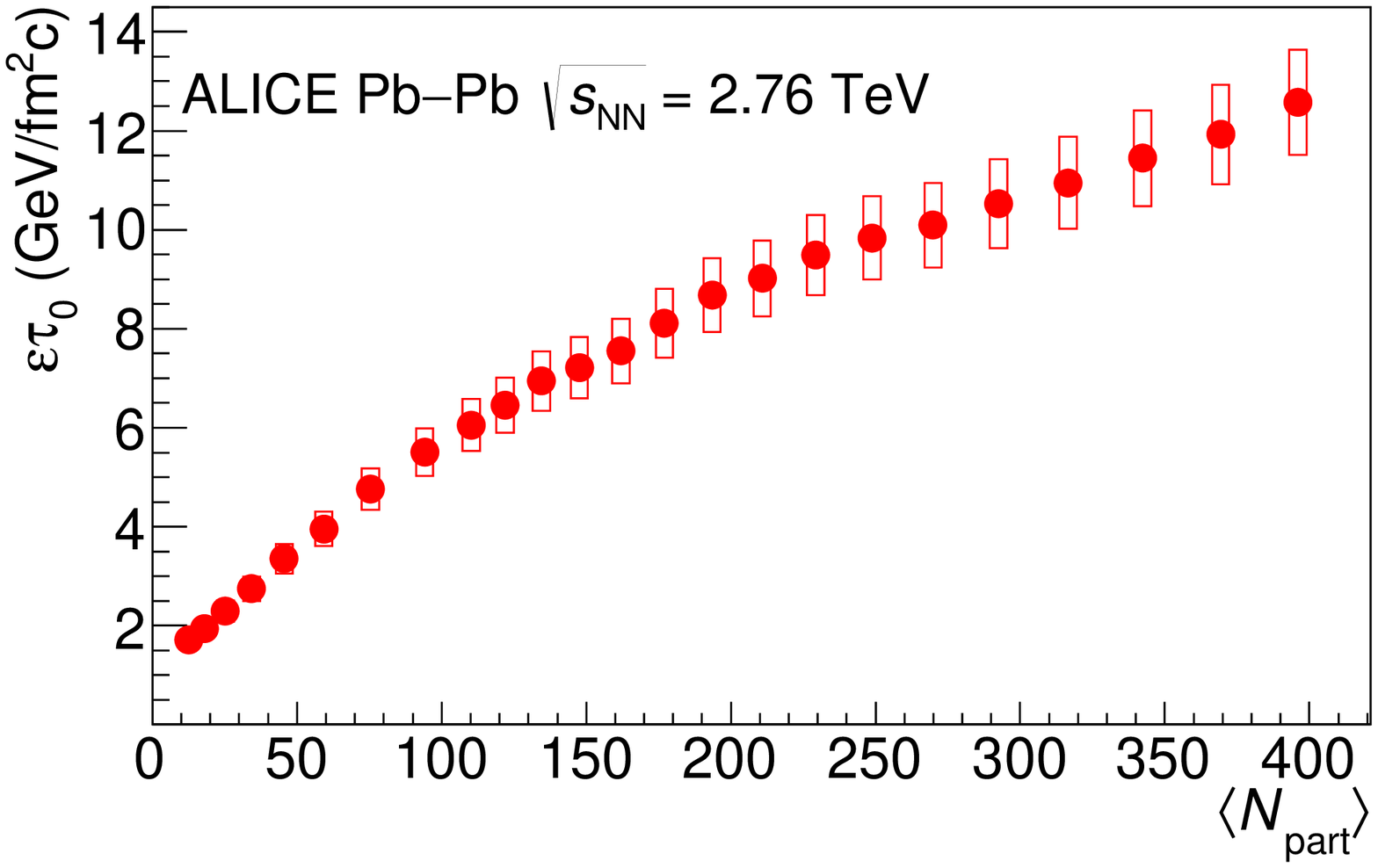}
\caption{$\epsilon\tau_{\mathrm 0}$ versus \Np estimated using \Eref{Eqtn:Bjorken}, $R = 7.17$ fm,  and the measured \meandEtdeta.  The boxes indicate the systematic uncertainties.}\label{Fig:EpsilonTauVsNpartWithModels}
\end{center}
\end{figure}

The transverse overlap area corresponding to the measured \meandEtdeta was determined by a calculation using a Glauber Monte Carlo method. Using the Glauber parameters from~\cite{Abelev:2013qoq} and assuming each participating nucleon has an effective transverse radius of $R = (\sigma_{\mathrm {NN}}^{\mathrm{inel}}/{4\pi})^{1/2} = 0.71$ fm results in $A = 162.5$ fm$^2$ for central collisions ($b=0$ fm). This is equivalent to a transverse overlap radius of $R=7.19$ fm, which is close to the value of 
7.17 fm often used in estimates of energy densities using a Woods-Saxon distribution to determine the effective area~\cite{Adler:2013aqf,Adare:2015bua}. The centrality dependence of A is obtained by assuming it scales as $(\sigma_{\mathrm x}^2 \sigma_{\mathrm y}^2 - \sigma_{\mathrm{xy}}^2)^{1/2}$~\cite{Alver:2008zza}, where $\sigma_{\mathrm x}^2$ and $\sigma_{\mathrm y}^2$ are the variances and $\sigma_{\mathrm{xy}}^2$ is the covariance of the spatial distribution of the participating nucleons in the transverse plane in the Glauber Monte Carlo calculation.   For 0--5\% central collisions this leads to a reduction of A by 3\% resulting in $\epsilon\tau_{0} = $ \energydensitytauE.
For comparison using  $R =7.17$ fm gives $\epsilon\tau_{0} = $ \energydensitytauB, roughly \energydensityfactor times that observed in 0--5\% central \Au collisions at \sNN = 200 GeV.  Some of this increase comes from the higher \npart in central \Pb collisions relative to central \Au collisions
The energy density times the formation time  $\epsilon\tau_{0}$ is shown in \Fref{Fig:EpsilonTauVsNpartWithModels} for $R$ = \RB, the same value of R used by PHENIX at RHIC energies~\cite{Adler:2013aqf,Adare:2015bua}.
%  (the same value of R used by Phenix at RHIC energies [28, 65])."

In addition to estimating the volume averaged energy density it is also interesting to estimate the energy density attained at the core of the collision area. This can be done by rewriting the Bjorken \Eref{Eqtn:Bjorken} as
\begin{equation}
 \epsilon_{\mathrm c} \tau_0 = \frac{\mathrm J}{c} \frac{\Big\langle \frac{{\mathrm d}E_{\mathrm T}}{{\mathrm d}\eta} \Big\rangle_{\mathrm c}}{ A_{\mathrm c} } 
                           = \frac{\mathrm J}{c} \frac{\Big\langle \frac{{\mathrm d}E_{\mathrm T}}{{\mathrm d}\eta} \Big\rangle}{\langle N_{\mathrm {part}} \rangle}  \sigma_{\mathrm {c} }
\label{Eqtn:Bjorken_2}
\end{equation}
where $A_{\mathrm c}$ is the area of the transverse core, $\Big\langle \frac{{\mathrm d}E_{\mathrm T}}{{\mathrm d}\eta} \Big\rangle_{\mathrm c}$ is $\Big\langle \frac{{\mathrm d}E_{\mathrm T}}{{\mathrm d}\eta} \Big\rangle$ produced in the core, and $ \sigma_{\mathrm {c}} =  \langle N_{\mathrm {part}}\rangle_{\mathrm c}/A_{\mathrm c}$ is the transverse area density of nucleon participants at the core.  The area $A_{\mathrm c}$ was chosen arbitrarily to be a circle with a radius of 1 fm at the center of the collision. Equation~\ref{Eqtn:Bjorken_2} assumes that the local energy density scales with the participant density in the transverse plane and that the measured value of $\Big\langle \frac{{\mathrm d}E_{\mathrm T}}{{\mathrm d}\eta} \Big\rangle / \langle N_{\mathrm {part}} \rangle$, which is averaged over the total transverse collision area, is also representative of the transverse energy production at the core, $\Big\langle \frac{{\mathrm d}E_{\mathrm T}}{{\mathrm d}\eta} \Big\rangle_{\mathrm c} / \langle N_{\mathrm {part}} \rangle_{\mathrm c}$. The increase of this quantity with increasing centrality indicates that this is a conservative estimate. From a Glauber Monte Carlo calculation we find for 0--5\% centrality $\sigma_{\mathrm c} = 4.2 \pm 0.1$ nucleon/fm$^2$ resulting in a core energy density of 
 $\epsilon_{\mathrm c} \tau_0 = 21 \pm 2$ GeV/fm$^2$/$c$.
For the most central 80 fm$^2$ (half the total overlap area) the energy density is still above 80\% of the core energy density emphasizing that the core energy density may be more relevant for judging the initial conditions of the QGP than the volume averaged energy density.

\section{Conclusions}\label{Sec:Conclusions}
\makeatletter{}We have measured \meandEtdeta at midrapidity in \Pb collisions at \sNN = 2.76 TeV using four different methods.  All methods lead to comparable results, although the systematic uncertainties are largely correlated.  Our results are consistent with results from CMS~\cite{CMSET:2012} for 10--80\% central collisions, however, we observe a lower \meandEtdeta in 0--10\% central collisions.  The \meandEtdeta observed at \sNN = 2.76 TeV in 0--5\% central collisions is \centralet.  The shape of the centrality dependence of \meandEtdetaPerNp is similar for RHIC and the LHC.   No centrality dependence of \meandEtdetaOverdNchdeta is observed within uncertainties, as was observed at RHIC.  Unlike at RHIC, we observe an increase in \meandEtdetaPerNq with centrality below \npart~$\approx$200.  Both \meandEtdetaPerNp and \meandEtdetaOverdNchdeta in central collisions exceed the value expected from naive extrapolations from data at lower collision energies.  Assuming that the formation time $\tau_{\mathrm 0}$ is 1 fm/$c$ the energy density is estimated to be at least \energydensity in 0--5\% central \Pb collisions at \sNN = 2.76 TeV and the energy density at the core of the collision exceeds 21 $\pm$ 2 GeV/fm$^3$.

\newenvironment{acknowledgement}{\relax}{\relax}
\begin{acknowledgement}
\section*{Acknowledgements}
% $Id: acknowledgements.tex 2287 2015-10-20 20:30:56Z loizides $
% Version: Nov 2015

The ALICE Collaboration would like to thank all its engineers and technicians for their invaluable contributions to the construction of the experiment and the CERN accelerator teams for the outstanding performance of the LHC complex.
The ALICE Collaboration gratefully acknowledges the resources and support provided by all Grid centres and the Worldwide LHC Computing Grid (WLCG) collaboration.
The ALICE Collaboration acknowledges the following funding agencies for their support in building and
running the ALICE detector:
State Committee of Science,  World Federation of Scientists (WFS)
and Swiss Fonds Kidagan, Armenia;
Conselho Nacional de Desenvolvimento Cient\'{\i}fico e Tecnol\'{o}gico (CNPq), Financiadora de Estudos e Projetos (FINEP),
Funda\c{c}\~{a}o de Amparo \`{a} Pesquisa do Estado de S\~{a}o Paulo (FAPESP);
National Natural Science Foundation of China (NSFC), the Chinese Ministry of Education (CMOE)
and the Ministry of Science and Technology of China (MSTC);
Ministry of Education and Youth of the Czech Republic;
Danish Natural Science Research Council, the Carlsberg Foundation and the Danish National Research Foundation;
The European Research Council under the European Community's Seventh Framework Programme;
Helsinki Institute of Physics and the Academy of Finland;
French CNRS-IN2P3, the `Region Pays de Loire', `Region Alsace', `Region Auvergne' and CEA, France;
German Bundesministerium fur Bildung, Wissenschaft, Forschung und Technologie (BMBF) and the Helmholtz Association;
General Secretariat for Research and Technology, Ministry of Development, Greece;
National Research, Development and Innovation Office (NKFIH), Hungary;
Department of Atomic Energy and Department of Science and Technology of the Government of India;
Istituto Nazionale di Fisica Nucleare (INFN) and Centro Fermi -
Museo Storico della Fisica e Centro Studi e Ricerche ``Enrico Fermi'', Italy;
Japan Society for the Promotion of Science (JSPS) KAKENHI and MEXT, Japan;
Joint Institute for Nuclear Research, Dubna;
National Research Foundation of Korea (NRF);
Consejo Nacional de Cienca y Tecnologia (CONACYT), Direccion General de Asuntos del Personal Academico(DGAPA), M\'{e}xico, Amerique Latine Formation academique - 
European Commission~(ALFA-EC) and the EPLANET Program~(European Particle Physics Latin American Network);
Stichting voor Fundamenteel Onderzoek der Materie (FOM) and the Nederlandse Organisatie voor Wetenschappelijk Onderzoek (NWO), Netherlands;
Research Council of Norway (NFR);
National Science Centre, Poland;
Ministry of National Education/Institute for Atomic Physics and National Council of Scientific Research in Higher Education~(CNCSI-UEFISCDI), Romania;
Ministry of Education and Science of Russian Federation, Russian
Academy of Sciences, Russian Federal Agency of Atomic Energy,
Russian Federal Agency for Science and Innovations and The Russian
Foundation for Basic Research;
Ministry of Education of Slovakia;
Department of Science and Technology, South Africa;
Centro de Investigaciones Energeticas, Medioambientales y Tecnologicas (CIEMAT), E-Infrastructure shared between Europe and Latin America (EELA), 
Ministerio de Econom\'{i}a y Competitividad (MINECO) of Spain, Xunta de Galicia (Conseller\'{\i}a de Educaci\'{o}n),
Centro de Aplicaciones Tecnológicas y Desarrollo Nuclear (CEA\-DEN), Cubaenerg\'{\i}a, Cuba, and IAEA (International Atomic Energy Agency);
Swedish Research Council (VR) and Knut $\&$ Alice Wallenberg
Foundation (KAW);
Ukraine Ministry of Education and Science;
United Kingdom Science and Technology Facilities Council (STFC);
The United States Department of Energy, the United States National
Science Foundation, the State of Texas, and the State of Ohio;
Ministry of Science, Education and Sports of Croatia and  Unity through Knowledge Fund, Croatia;
Council of Scientific and Industrial Research (CSIR), New Delhi, India;
Pontificia Universidad Cat\'{o}lica del Per\'{u}.
    %%%%%%% get the lates version before submitting
\end{acknowledgement}

\bibliographystyle{utphys}   
\bibliography{Bibliography}   

\providecommand{\href}[2]{#2}\begingroup\raggedright\begin{thebibliography}{10}

\bibitem{Karch:2002abc}
F.~Karsch, \href{http://dx.doi.org/10.1007/3-540-45792-5_6}{``Lattice qcd at
  high temperature and density,''} in {\em Lectures on Quark Matter},
  W.~Plessas and L.~Mathelitsch, eds., vol.~583 of {\em Lecture Notes in
  Physics}, pp.~209--249.
\newblock Springer Berlin Heidelberg, 2002.
\newblock \url{http://dx.doi.org/10.1007/3-540-45792-5_6}.

\bibitem{Bazavov:2014pvz}
{\bfseries HotQCD} Collaboration, A.~Bazavov {\em et~al.}, ``{Equation of state
  in ( 2+1 )-flavor QCD},''
  \href{http://dx.doi.org/10.1103/PhysRevD.90.094503}{{\em Phys. Rev.}
  {\bfseries D90} no.~9, (2014) 094503},
\href{http://arxiv.org/abs/1407.6387}{{\ttfamily arXiv:1407.6387 [hep-lat]}}.
%%CITATION = ARXIV:1407.6387;%%.

\bibitem{Adcox:2004mh}
{\bfseries PHENIX} Collaboration, K.~Adcox {\em et~al.}, ``{Formation of dense
  partonic matter in relativistic nucleus nucleus collisions at RHIC:
  Experimental evaluation by the PHENIX collaboration},''
  \href{http://dx.doi.org/10.1016/j.nuclphysa.2005.03.086}{{\em Nucl. Phys.}
  {\bfseries A757} (2005) 184--283},
\href{http://arxiv.org/abs/nucl-ex/0410003}{{\ttfamily arXiv:nucl-ex/0410003}}.
%%CITATION = NUCL-EX/0410003;%%.

\bibitem{Adams:2005dq}
{\bfseries STAR} Collaboration, J.~Adams {\em et~al.}, ``{Experimental and
  theoretical challenges in the search for the quark gluon plasma: The STAR
  collaboration's critical assessment of the evidence from RHIC collisions},''
  \href{http://dx.doi.org/10.1016/j.nuclphysa.2005.03.085}{{\em Nucl. Phys.}
  {\bfseries A757} (2005) 102--183},
\href{http://arxiv.org/abs/nucl-ex/0501009}{{\ttfamily arXiv:nucl-ex/0501009}}.
%%CITATION = NUCL-EX/0501009;%%.

\bibitem{Back:2004je}
{\bfseries PHOBOS} Collaboration, B.~B. Back {\em et~al.}, ``{The PHOBOS
  perspective on discoveries at RHIC},''
  \href{http://dx.doi.org/10.1016/j.nuclphysa.2005.03.084}{{\em Nucl. Phys.}
  {\bfseries A757} (2005) 28--101},
\href{http://arxiv.org/abs/nucl-ex/0410022}{{\ttfamily arXiv:nucl-ex/0410022}}.
%%CITATION = NUCL-EX/0410022;%%.

\bibitem{Arsene:2004fa}
{\bfseries BRAHMS} Collaboration, I.~Arsene {\em et~al.}, ``{Quark Gluon Plasma
  and Color Glass Condensate at RHIC? The perspective from the BRAHMS
  experiment},'' \href{http://dx.doi.org/10.1016/j.nuclphysa.2005.02.130}{{\em
  Nucl. Phys.} {\bfseries A757} (2005) 1--27},
\href{http://arxiv.org/abs/nucl-ex/0410020}{{\ttfamily arXiv:nucl-ex/0410020}}.
%%CITATION = NUCL-EX/0410020;%%.

\bibitem{Aamodt:2011mr}
{\bfseries ALICE} Collaboration, K.~Aamodt {\em et~al.}, ``{Two-pion
  Bose-Einstein correlations in central Pb-Pb collisions at $\sqrt{{s}_{NN}} =$
  2.76 TeV},'' \href{http://dx.doi.org/10.1016/j.physletb.2010.12.053}{{\em
  Phys. Lett.} {\bfseries B696} (2011) 328--337},
\href{http://arxiv.org/abs/1012.4035}{{\ttfamily arXiv:1012.4035 [nucl-ex]}}.
%%CITATION = ARXIV:1012.4035;%%.

\bibitem{Aamodt:2010cz}
{\bfseries ALICE} Collaboration, K.~Aamodt {\em et~al.}, ``{Centrality
  dependence of the charged-particle multiplicity density at mid-rapidity in
  Pb-Pb collisions at \sNN = 2.76 TeV},'' {\em Phys. Rev. Lett.} {\bfseries
  106} (2011) 032301,
\href{http://arxiv.org/abs/1012.1657}{{\ttfamily arXiv:1012.1657 [nucl-ex]}}.
%%CITATION = 1012.1657;%%.

\bibitem{Aamodt:2010jd}
{\bfseries ALICE} Collaboration, K.~Aamodt {\em et~al.}, ``{Suppression of
  Charged Particle Production at Large Transverse Momentum in Central Pb-Pb
  Collisions at $\sqrt{s_{NN}} =$ 2.76 TeV},''
  \href{http://dx.doi.org/10.1016/j.physletb.2010.12.020}{{\em Phys. Lett.}
  {\bfseries B696} (2011) 30--39},
\href{http://arxiv.org/abs/1012.1004}{{\ttfamily arXiv:1012.1004 [nucl-ex]}}.
%%CITATION = ARXIV:1012.1004;%%.

\bibitem{Aamodt:2010pb}
{\bfseries ALICE} Collaboration, K.~Aamodt {\em et~al.}, ``{Charged-particle
  multiplicity density at mid-rapidity in central Pb-Pb collisions at \sNN =
  2.76 TeV},'' {\em Phys. Rev. Lett.} {\bfseries 105} (2010) 252301,
\href{http://arxiv.org/abs/1011.3916}{{\ttfamily arXiv:1011.3916 [nucl-ex]}}.
%%CITATION = 1011.3916;%%.

\bibitem{Aamodt:2010pa}
{\bfseries ALICE} Collaboration, K.~Aamodt {\em et~al.}, ``{Elliptic flow of
  charged particles in Pb-Pb collisions at 2.76 TeV},''
  \href{http://dx.doi.org/10.1103/PhysRevLett.105.252302}{{\em Phys. Rev.
  Lett.} {\bfseries 105} (2010) 252302},
\href{http://arxiv.org/abs/1011.3914}{{\ttfamily arXiv:1011.3914 [nucl-ex]}}.
%%CITATION = ARXIV:1011.3914;%%.

\bibitem{Aad:2010bu}
{\bfseries ATLAS} Collaboration, G.~Aad {\em et~al.}, ``{Observation of a
  Centrality-Dependent Dijet Asymmetry in Lead-Lead Collisions at
  $\sqrt{s_{NN}}=2.76$ TeV with the ATLAS Detector at the LHC},''
  \href{http://dx.doi.org/10.1103/PhysRevLett.105.252303}{{\em Phys. Rev.
  Lett.} {\bfseries 105} (2010) 252303},
\href{http://arxiv.org/abs/1011.6182}{{\ttfamily arXiv:1011.6182 [hep-ex]}}.
%%CITATION = ARXIV:1011.6182;%%.

\bibitem{Chatrchyan:2011sx}
{\bfseries CMS} Collaboration, S.~Chatrchyan {\em et~al.}, ``{Observation and
  studies of jet quenching in PbPb collisions at nucleon-nucleon center-of-mass
  energy = 2.76 TeV},''
  \href{http://dx.doi.org/10.1103/PhysRevC.84.024906}{{\em Phys. Rev.}
  {\bfseries C84} (2011) 024906},
\href{http://arxiv.org/abs/1102.1957}{{\ttfamily arXiv:1102.1957 [nucl-ex]}}.
%%CITATION = ARXIV:1102.1957;%%.

\bibitem{Chatrchyan:2011pb}
{\bfseries CMS} Collaboration, S.~Chatrchyan {\em et~al.}, ``{Dependence on
  pseudorapidity and centrality of charged hadron production in PbPb collisions
  at a nucleon-nucleon centre-of-mass energy of 2.76 TeV},''
  \href{http://dx.doi.org/10.1007/JHEP08(2011)141}{{\em JHEP} {\bfseries 08}
  (2011) 141},
\href{http://arxiv.org/abs/1107.4800}{{\ttfamily arXiv:1107.4800 [nucl-ex]}}.
%%CITATION = ARXIV:1107.4800;%%.

\bibitem{Chatrchyan:2012ta}
{\bfseries CMS} Collaboration, S.~Chatrchyan {\em et~al.}, ``{Measurement of
  the elliptic anisotropy of charged particles produced in PbPb collisions at
  $\sqrt{s}_{NN}$=2.76 TeV},''
  \href{http://dx.doi.org/10.1103/PhysRevC.87.014902}{{\em Phys. Rev.}
  {\bfseries C87} no.~1, (2013) 014902},
\href{http://arxiv.org/abs/1204.1409}{{\ttfamily arXiv:1204.1409 [nucl-ex]}}.
%%CITATION = ARXIV:1204.1409;%%.

\bibitem{ATLAS:2011ag}
{\bfseries ATLAS} Collaboration, G.~Aad {\em et~al.}, ``{Measurement of the
  centrality dependence of the charged particle pseudorapidity distribution in
  lead-lead collisions at $\sqrt{s_{NN}}=2.76$ TeV with the ATLAS detector},''
  \href{http://dx.doi.org/10.1016/j.physletb.2012.02.045}{{\em Phys. Lett.}
  {\bfseries B710} (2012) 363--382},
\href{http://arxiv.org/abs/1108.6027}{{\ttfamily arXiv:1108.6027 [hep-ex]}}.
%%CITATION = ARXIV:1108.6027;%%.

\bibitem{ATLAS:2011ah}
{\bfseries ATLAS} Collaboration, G.~Aad {\em et~al.}, ``{Measurement of the
  pseudorapidity and transverse momentum dependence of the elliptic flow of
  charged particles in lead-lead collisions at $\sqrt{s_{NN}}=2.76$ TeV with
  the ATLAS detector},''
  \href{http://dx.doi.org/10.1016/j.physletb.2011.12.056}{{\em Phys. Lett.}
  {\bfseries B707} (2012) 330--348},
\href{http://arxiv.org/abs/1108.6018}{{\ttfamily arXiv:1108.6018 [hep-ex]}}.
%%CITATION = ARXIV:1108.6018;%%.

\bibitem{Bjorken:1982qr}
J.~D. Bjorken, ``{Highly Relativistic Nucleus-Nucleus Collisions: The Central
  Rapidity Region},''
\href{http://dx.doi.org/10.1103/PhysRevD.27.140}{{\em Phys. Rev.} {\bfseries
  D27} (1983) 140--151}.
%%CITATION = PHRVA,D27,140;%%.

\bibitem{Ahle:1994nk}
{\bfseries E-802} Collaboration, L.~Ahle {\em et~al.}, ``{Global transverse
  energy distributions in Si + Al, Au at 14.6-A/GeV/c and Au + Au at
  11.6-A.GeV/c},''
\href{http://dx.doi.org/10.1016/0370-2693(94)91251-3}{{\em Phys. Lett.}
  {\bfseries B332} (1994) 258--264}.
%%CITATION = PHLTA,B332,258;%%.

\bibitem{Barrette:1993pm}
{\bfseries E814/E877} Collaboration, J.~Barrette {\em et~al.}, ``{Measurement
  of transverse energy production with Si and Au beams at relativistic energy:
  Towards hot and dense hadronic matter},''
\href{http://dx.doi.org/10.1103/PhysRevLett.70.2996}{{\em Phys. Rev. Lett.}
  {\bfseries 70} (1993) 2996--2999}.
%%CITATION = PRLTA,70,2996;%%.

\bibitem{Akesson:1987kh}
{\bfseries HELIOS} Collaboration, T.~Akesson {\em et~al.}, ``{The Transverse
  Energy Distribution in $^{16}$O - Nucleus Collisions at 60-{GeV} and
  200-{GeV} Per Nucleon},''
{\em Z. Phys.} {\bfseries C38} (1988) 383.
%%CITATION = ZEPYA,C38,383;%%.

\bibitem{Bamberger:1986mt}
{\bfseries NA35} Collaboration, A.~Bamberger {\em et~al.}, ``{Multiplicity and
  Transverse Energy Flux in $^{16}$O Pb at 200-{GeV} Per Nucleon},''
\href{http://dx.doi.org/10.1016/0370-2693(87)90581-8}{{\em Phys. Lett.}
  {\bfseries B184} (1987) 271}.
%%CITATION = PHLTA,B184,271;%%.

\bibitem{Margetis:1994tt}
{\bfseries NA49} Collaboration, T.~Alber {\em et~al.}, ``{Transverse energy
  production in Pb-208 + Pb collisions at 158-GeV per nucleon},''
\href{http://dx.doi.org/10.1103/PhysRevLett.75.3814}{{\em Phys. Rev. Lett.}
  {\bfseries 75} (1995) 3814--3817}.
%%CITATION = PRLTA,75,3814;%%.

\bibitem{WA80Et:1991}
{\bfseries WA80} Collaboration, R.~Albrecht {\em et~al.}, ``{Distributions of
  transverse energy and forward energy in O- and S-induced heavy ion collisions
  at 60A and 200A Gev},'' {\em Phys. Rev.} {\bfseries C44} (1991) 2736--2752.

\bibitem{Aggarwal:2000bc}
{\bfseries WA98} Collaboration, M.~Aggarwal {\em et~al.}, ``{Scaling of
  particle and transverse energy production in Pb-208 + Pb-208 collisions at
  158-A-GeV},'' \href{http://dx.doi.org/10.1007/s100520100578}{{\em Eur. Phys.
  J.} {\bfseries C18} (2001) 651--663},
\href{http://arxiv.org/abs/nucl-ex/0008004}{{\ttfamily arXiv:nucl-ex/0008004
  [nucl-ex]}}.
%%CITATION = NUCL-EX/0008004;%%.

\bibitem{Adler:2004zn}
{\bfseries PHENIX} Collaboration, S.~Adler {\em et~al.}, ``{Systematic studies
  of the centrality and s(NN)**(1/2) dependence of the d E(T) / d eta and d
  (N(ch) / d eta in heavy ion collisions at mid-rapidity},''
  \href{http://dx.doi.org/10.1103/PhysRevC.71.049901,
  10.1103/PhysRevC.71.034908, 10.1103/PhysRevC.71.034908
  10.1103/PhysRevC.71.049901}{{\em Phys. Rev.} {\bfseries C71} (2005) 034908},
\href{http://arxiv.org/abs/nucl-ex/0409015}{{\ttfamily arXiv:nucl-ex/0409015
  [nucl-ex]}}.
%%CITATION = NUCL-EX/0409015;%%.

\bibitem{Adcox:2001ry}
{\bfseries PHENIX} Collaboration, K.~Adcox {\em et~al.}, ``{Measurement of the
  mid-rapidity transverse energy distribution from $\sqrt{s_{NN}}$ = 130-GeV Au
  + Au collisions at RHIC},''
  \href{http://dx.doi.org/10.1103/PhysRevLett.87.052301}{{\em Phys. Rev. Lett.}
  {\bfseries 87} (2001) 052301},
\href{http://arxiv.org/abs/nucl-ex/0104015}{{\ttfamily arXiv:nucl-ex/0104015}}.
%%CITATION = NUCL-EX/0104015;%%.

\bibitem{Adler:2013aqf}
{\bfseries PHENIX} Collaboration, S.~Adler {\em et~al.}, ``{Transverse-energy
  distributions at midrapidity in p+p , d+Au , and Au+Au collisions at
  $\sqrt{s_{NN}}=62.4-€$ GeV and implications for particle-production
  models},'' \href{http://dx.doi.org/10.1103/PhysRevC.89.044905}{{\em Phys.
  Rev.} {\bfseries C89} no.~4, (2014) 044905},
\href{http://arxiv.org/abs/1312.6676}{{\ttfamily arXiv:1312.6676 [nucl-ex]}}.
%%CITATION = ARXIV:1312.6676;%%.

\bibitem{Adams:2004cb}
{\bfseries STAR} Collaboration, J.~Adams {\em et~al.}, ``{Measurements of
  transverse energy distributions in Au + Au collisions at \sNN = 200-GeV},''
  \href{http://dx.doi.org/10.1103/PhysRevC.70.054907}{{\em Phys. Rev.}
  {\bfseries C70} (2004) 054907},
\href{http://arxiv.org/abs/nucl-ex/0407003}{{\ttfamily arXiv:nucl-ex/0407003}}.
%%CITATION = NUCL-EX/0407003;%%.

\bibitem{CMSET:2012}
{\bfseries CMS} Collaboration, S.~Chatrchyan {\em et~al.}, ``{Measurement of
  the Pseudorapidity and Centrality Dependence of the Transverse Energy Density
  in Pb-Pb Collisions at \sqrts = 2.76 TeV},'' {\em Phys. Rev. Lett.}
  {\bfseries 109} (2012) 152303.

\bibitem{Bialas:1976ed}
A.~Bialas, M.~Bleszynski, and W.~Czyz, ``{Multiplicity Distributions in
  Nucleus-Nucleus Collisions at High-Energies},''
\href{http://dx.doi.org/10.1016/0550-3213(76)90329-1}{{\em Nucl. Phys.}
  {\bfseries B111} (1976) 461}.
%%CITATION = NUPHA,B111,461;%%.

\bibitem{Miller:2007}
M.~Miller, K.~Reygers, S.~Sanders, and P.~Steinberg, ``{Glauber Modeling in
  High-Energy Nuclear Collisions},'' {\em Annu. Rev. Nucl. Part. Sci.}
  {\bfseries 57} (2007) 205--243.

\bibitem{Wang:2000bf}
X.-N. Wang and M.~Gyulassy, ``{Energy and centrality dependence of rapidity
  densities at RHIC},''
  \href{http://dx.doi.org/10.1103/PhysRevLett.86.3496}{{\em Phys. Rev. Lett.}
  {\bfseries 86} (2001) 3496--3499},
\href{http://arxiv.org/abs/nucl-th/0008014}{{\ttfamily arXiv:nucl-th/0008014
  [nucl-th]}}.
%%CITATION = NUCL-TH/0008014;%%.

\bibitem{Trainor:2015ida}
T.~A. Trainor, ``{Questioning tests of hadron production models based on
  minimum-bias distributions on transverse energy},''
\href{http://dx.doi.org/10.1103/PhysRevC.91.044905}{{\em Phys. Rev.} {\bfseries
  C91} no.~4, (2015) 044905}.
%%CITATION = PHRVA,C91,044905;%%.

\bibitem{Eremin:2003}
S.~Eremin and S.~Voloshin, ``{Nucleon Participants or quark participants?},''
  {\em Phys. Rev.} {\bfseries C67} (2003) 064905.

\bibitem{Adler:2014a}
{\bfseries PHENIX} Collaboration, S.~Adler {\em et~al.}, ``{Transverse-energy
  distributions at midrapidity in p+p, d+Au, and Au+Au collisions at \sqrts =
  62.4–200 GeV and implications for particle-production models},'' {\em Phys.
  Rev.} {\bfseries C89} (2014) 044905.

\bibitem{Aamodt:2008zz}
{\bfseries ALICE} Collaboration, K.~Aamodt {\em et~al.}, ``{The ALICE
  experiment at the CERN LHC},''
\href{http://dx.doi.org/10.1088/1748-0221/3/08/S08002}{{\em JINST} {\bfseries
  3} (2008) S08002}.
%%CITATION = JINST,3,S08002;%%.

\bibitem{Abbas:2013taa}
{\bfseries ALICE} Collaboration, E.~Abbas {\em et~al.}, ``{Performance of the
  ALICE VZERO system},''
  \href{http://dx.doi.org/10.1088/1748-0221/8/10/P10016}{{\em JINST} {\bfseries
  8} (2013) P10016},
\href{http://arxiv.org/abs/1306.3130}{{\ttfamily arXiv:1306.3130 [nucl-ex]}}.
%%CITATION = ARXIV:1306.3130;%%.

\bibitem{Alme:2010ke}
J.~Alme {\em et~al.}, ``{The ALICE TPC, a large 3-dimensional tracking device
  with fast readout for ultra-high multiplicity events},''
  \href{http://dx.doi.org/10.1016/j.nima.2010.04.042}{{\em Nucl. Instrum.
  Meth.} {\bfseries A622} (2010) 316--367},
\href{http://arxiv.org/abs/1001.1950}{{\ttfamily arXiv:1001.1950
  [physics.ins-det]}}.
%%CITATION = ARXIV:1001.1950;%%.

\bibitem{Aamodt:2010aa}
{\bfseries ALICE} Collaboration, K.~Aamodt {\em et~al.}, ``{Alignment of the
  ALICE Inner Tracking System with cosmic-ray tracks},''
  \href{http://dx.doi.org/10.1088/1748-0221/5/03/P03003}{{\em JINST} {\bfseries
  5} (2010) P03003},
\href{http://arxiv.org/abs/1001.0502}{{\ttfamily arXiv:1001.0502
  [physics.ins-det]}}.
%%CITATION = ARXIV:1001.0502;%%.

\bibitem{Allen:2009aa}
{\bfseries ALICE EMCal} Collaboration, J.~Allen {\em et~al.}, ``{Performance of
  prototypes for the ALICE electromagnetic calorimeter},''
  \href{http://dx.doi.org/10.1016/j.nima.2009.12.061}{{\em Nucl. Instrum.
  Meth.} {\bfseries A615} (2010) 6--13},
\href{http://arxiv.org/abs/0912.2005}{{\ttfamily arXiv:0912.2005
  [physics.ins-det]}}.
%%CITATION = ARXIV:0912.2005;%%.

\bibitem{Cortese:2008zza}
{\bfseries ALICE} Collaboration, P.~Cortese {\em et~al.},
``{ALICE electromagnetic calorimeter technical design report},''.
%%CITATION = CERN-LHCC-2008-014 ETC.;%%.

\bibitem{Dellacasa:1999kd}
{\bfseries ALICE} Collaboration, G.~Dellacasa {\em et~al.},
``{ALICE technical design report of the photon spectrometer (PHOS)},''.
%%CITATION = CERN-LHCC-99-04 ETC.;%%.

\bibitem{Aleksandrov:2005yu}
{\bfseries ALICE PHOS calorimeter} Collaboration, D.~V. Aleksandrov {\em
  et~al.}, ``{A high resolution electromagnetic calorimeter based on
  lead-tungstate crystals},''
\href{http://dx.doi.org/10.1016/j.nima.2005.03.174}{{\em Nucl. Instrum. Meth.}
  {\bfseries A550} (2005) 169--184}.
%%CITATION = NUIMA,A550,169;%%.

\bibitem{Abelev:2014ffa}
{\bfseries ALICE} Collaboration, B.~B. Abelev {\em et~al.}, ``{Performance of
  the ALICE Experiment at the CERN LHC},''
  \href{http://dx.doi.org/10.1142/S0217751X14300440}{{\em Int. J. Mod. Phys.}
  {\bfseries A29} (2014) 1430044},
\href{http://arxiv.org/abs/1402.4476}{{\ttfamily arXiv:1402.4476 [nucl-ex]}}.
%%CITATION = ARXIV:1402.4476;%%.

\bibitem{Abelev:2013vea}
{\bfseries ALICE} Collaboration, B.~Abelev {\em et~al.}, ``{Centrality
  dependence of $\pi$, K, p production in Pb-Pb collisions at $\sqrt{s_{NN}}$ =
  2.76 TeV},'' \href{http://dx.doi.org/10.1103/PhysRevC.88.044910}{{\em Phys.
  Rev.} {\bfseries C88} (2013) 044910},
\href{http://arxiv.org/abs/1303.0737}{{\ttfamily arXiv:1303.0737 [hep-ex]}}.
%%CITATION = ARXIV:1303.0737;%%.

\bibitem{Wang:1991hta}
X.-N. Wang and M.~Gyulassy, ``{HIJING: A Monte Carlo model for multiple jet
  production in p p, p A and A A collisions},''
\href{http://dx.doi.org/10.1103/PhysRevD.44.3501}{{\em Phys. Rev.} {\bfseries
  D44} (1991) 3501--3516}.
%%CITATION = PHRVA,D44,3501;%%.

\bibitem{Brun:1994aa}
R.~Brun, F.~Carminati, and S.~Giani,
``{GEANT Detector Description and Simulation Tool},''.
%%CITATION = CERN-W5013 ETC.;%%.

\bibitem{Abelev:2014uua}
{\bfseries ALICE} Collaboration, B.~B. Abelev {\em et~al.}, ``{K*(892) and
  phi(1020) production in Pb-Pb collisions at \sNN=2.76 TeV},''
  \href{http://dx.doi.org/10.1103/PhysRevC.91.024609}{{\em Phys. Rev.}
  {\bfseries C91} (2015) 024609},
\href{http://arxiv.org/abs/1404.0495}{{\ttfamily arXiv:1404.0495 [nucl-ex]}}.
%%CITATION = ARXIV:1404.0495;%%.

\bibitem{Abelev:2013xaa}
{\bfseries ALICE} Collaboration, B.~B. Abelev {\em et~al.}, ``{$K^0_S$ and
  $\Lambda$ production in Pb-Pb collisions at $\sqrt{s_{NN}}$ = 2.76 TeV},''
  \href{http://dx.doi.org/10.1103/PhysRevLett.111.222301}{{\em Phys. Rev.
  Lett.} {\bfseries 111} (2013) 222301},
\href{http://arxiv.org/abs/1307.5530}{{\ttfamily arXiv:1307.5530 [nucl-ex]}}.
%%CITATION = ARXIV:1307.5530;%%.

\bibitem{Abelev:2013bla}
{\bfseries ALICE} Collaboration, B.~B. Abelev {\em et~al.}, ``{Multiplicity
  dependence of the average transverse momentum in pp, p-Pb, and Pb-Pb
  collisions at the LHC},''
  \href{http://dx.doi.org/10.1016/j.physletb.2013.10.054}{{\em Phys. Lett.}
  {\bfseries B727} (2013) 371--380},
\href{http://arxiv.org/abs/1307.1094}{{\ttfamily arXiv:1307.1094 [nucl-ex]}}.
%%CITATION = ARXIV:1307.1094;%%.

\bibitem{Sjostrand:2006za}
T.~Sjostrand, S.~Mrenna, and P.~Z. Skands, ``{PYTHIA 6.4 Physics and Manual},''
  \href{http://dx.doi.org/10.1088/1126-6708/2006/05/026}{{\em JHEP} {\bfseries
  0605} (2006) 026},
\href{http://arxiv.org/abs/hep-ph/0603175}{{\ttfamily arXiv:hep-ph/0603175
  [hep-ph]}}.
%%CITATION = HEP-PH/0603175;%%.

\bibitem{Wheaton:2004qb}
S.~Wheaton and J.~Cleymans, ``{THERMUS: A Thermal model package for ROOT},''
  \href{http://dx.doi.org/10.1016/j.cpc.2008.08.001}{{\em Comput. Phys.
  Commun.} {\bfseries 180} (2009) 84--106},
\href{http://arxiv.org/abs/hep-ph/0407174}{{\ttfamily arXiv:hep-ph/0407174
  [hep-ph]}}.
%%CITATION = HEP-PH/0407174;%%.

\bibitem{Adare:2010fe}
{\bfseries PHENIX} Collaboration, A.~Adare {\em et~al.}, ``{Measurement of
  neutral mesons in p+p collisions at $\sqrt(s)$= 200 GeV and scaling
  properties of hadron production},''
  \href{http://dx.doi.org/10.1103/PhysRevD.83.052004}{{\em Phys. Rev.}
  {\bfseries D83} (2011) 052004},
\href{http://arxiv.org/abs/1005.3674}{{\ttfamily arXiv:1005.3674 [hep-ex]}}.
%%CITATION = ARXIV:1005.3674;%%.

\bibitem{Adare:2010dc}
{\bfseries PHENIX} Collaboration, A.~Adare {\em et~al.}, ``{Transverse momentum
  dependence of meson suppression $\eta$ suppression in Au+Au collisions at
  $\sqrt{s_{NN}}$ = 200 GeV},''
  \href{http://dx.doi.org/10.1103/PhysRevC.82.011902}{{\em Phys. Rev.}
  {\bfseries C82} (2010) 011902},
\href{http://arxiv.org/abs/1005.4916}{{\ttfamily arXiv:1005.4916 [nucl-ex]}}.
%%CITATION = ARXIV:1005.4916;%%.

\bibitem{Abelev:2012cn}
{\bfseries ALICE} Collaboration, B.~Abelev {\em et~al.}, ``{Neutral pion and
  $\eta$ meson production in proton-proton collisions at $\sqrt{s}=0.9$ TeV and
  $\sqrt{s}=7$ TeV},''
  \href{http://dx.doi.org/10.1016/j.physletb.2012.09.015}{{\em Phys. Lett.}
  {\bfseries B717} (2012) 162--172},
\href{http://arxiv.org/abs/1205.5724}{{\ttfamily arXiv:1205.5724 [hep-ex]}}.
%%CITATION = ARXIV:1205.5724;%%.

\bibitem{PhysRevLett.105.072002}
{\bfseries ALICE} Collaboration, K.~Aamodt {\em et~al.}, ``Midrapidity
  antiproton-to-proton ratio in $pp$ collisons at $\sqrt{s}=0.9$ and 7~tev
  measured by the alice experiment,''
  \href{http://dx.doi.org/10.1103/PhysRevLett.105.072002}{{\em Phys. Rev.
  Lett.} {\bfseries 105} (Aug, 2010) 072002}.
  \url{http://link.aps.org/doi/10.1103/PhysRevLett.105.072002}.

\bibitem{Abbas:2013rua}
{\bfseries ALICE} Collaboration, E.~Abbas {\em et~al.}, ``{Mid-rapidity
  anti-baryon to baryon ratios in pp collisions at $\sqrt{s}$ = 0.9, 2.76 and 7
  TeV measured by ALICE},''
  \href{http://dx.doi.org/10.1140/epjc/s10052-013-2496-5}{{\em Eur. Phys. J.}
  {\bfseries C73} (2013) 2496},
\href{http://arxiv.org/abs/1305.1562}{{\ttfamily arXiv:1305.1562 [nucl-ex]}}.
%%CITATION = ARXIV:1305.1562;%%.

\bibitem{Abelev:2013qoq}
{\bfseries ALICE} Collaboration, B.~Abelev {\em et~al.}, ``{Centrality
  determination of Pb-Pb collisions at $\sqrt{s_{NN}}$ = 2.76 TeV with
  ALICE},'' \href{http://dx.doi.org/10.1103/PhysRevC.88.044909}{{\em Phys.
  Rev.} {\bfseries C88} no.~4, (2013) 044909},
\href{http://arxiv.org/abs/1301.4361}{{\ttfamily arXiv:1301.4361 [nucl-ex]}}.
%%CITATION = ARXIV:1301.4361;%%.

\bibitem{Abelev:2013fn}
{\bfseries ALICE} Collaboration, B.~Abelev {\em et~al.}, ``{Measurement of the
  inclusive differential jet cross section in $pp$ collisions at $\sqrt{s} =
  2.76$ TeV},'' \href{http://dx.doi.org/10.1016/j.physletb.2013.04.026}{{\em
  Phys. Lett.} {\bfseries B722} (2013) 262--272},
\href{http://arxiv.org/abs/1301.3475}{{\ttfamily arXiv:1301.3475 [nucl-ex]}}.
%%CITATION = ARXIV:1301.3475;%%.

\bibitem{Abelev:2014ypa}
{\bfseries ALICE} Collaboration, B.~B. Abelev {\em et~al.}, ``{Neutral pion
  production at midrapidity in pp and Pb-Pb collisions at $\sqrt{s_{{\mathrm
  {NN}}}}= 2.76\,{\mathrm {TeV}}$},''
  \href{http://dx.doi.org/10.1140/epjc/s10052-014-3108-8}{{\em Eur. Phys. J.}
  {\bfseries C74} no.~10, (2014) 3108},
\href{http://arxiv.org/abs/1405.3794}{{\ttfamily arXiv:1405.3794 [nucl-ex]}}.
%%CITATION = ARXIV:1405.3794;%%.

\bibitem{Loizides:2016djv}
C.~Loizides, ``{Glauber modeling of high-energy nuclear collisions at
  sub-nucleon level},''
\href{http://arxiv.org/abs/1603.07375}{{\ttfamily arXiv:1603.07375 [nucl-ex]}}.
%%CITATION = ARXIV:1603.07375;%%.

\bibitem{Abelev:2008ab}
{\bfseries STAR} Collaboration, B.~Abelev {\em et~al.}, ``{Systematic
  Measurements of Identified Particle Spectra in $p p, d^+$ Au and Au+Au
  Collisions from STAR},''
  \href{http://dx.doi.org/10.1103/PhysRevC.79.034909}{{\em Phys. Rev.}
  {\bfseries C79} (2009) 034909},
\href{http://arxiv.org/abs/0808.2041}{{\ttfamily arXiv:0808.2041 [nucl-ex]}}.
%%CITATION = ARXIV:0808.2041;%%.

\bibitem{Adler:2003cb}
{\bfseries PHENIX} Collaboration, S.~S. Adler {\em et~al.}, ``{Identified
  charged particle spectra and yields in Au + Au collisions at \sNN =
  200-GeV},'' \href{http://dx.doi.org/10.1103/PhysRevC.69.034909}{{\em Phys.
  Rev.} {\bfseries C69} (2004) 034909},
\href{http://arxiv.org/abs/nucl-ex/0307022}{{\ttfamily arXiv:nucl-ex/0307022}}.
%%CITATION = NUCL-EX/0307022;%%.

\bibitem{Adare:2015bua}
{\bfseries PHENIX} Collaboration, A.~Adare {\em et~al.}, ``{Transverse energy
  production and charged-particle multiplicity at midrapidity in various
  systems from $\sqrt{s_{NN}}=7.7$ to 200 GeV},''
\href{http://arxiv.org/abs/1509.06727}{{\ttfamily arXiv:1509.06727 [nucl-ex]}}.
%%CITATION = ARXIV:1509.06727;%%.

\bibitem{Eskola:1999fc}
K.~Eskola, K.~Kajantie, P.~Ruuskanen, and K.~Tuominen, ``{Scaling of transverse
  energies and multiplicities with atomic number and energy in
  ultrarelativistic nuclear collisions},''
  \href{http://dx.doi.org/10.1016/S0550-3213(99)00720-8}{{\em Nucl. Phys.}
  {\bfseries B570} (2000) 379--389},
\href{http://arxiv.org/abs/hep-ph/9909456}{{\ttfamily arXiv:hep-ph/9909456
  [hep-ph]}}.
%%CITATION = HEP-PH/9909456;%%.

\bibitem{Renk:2011gj}
T.~Renk, H.~Holopainen, R.~Paatelainen, and K.~J. Eskola, ``{Systematics of the
  charged-hadron $p_T$ spectrum and the nuclear suppression factor in heavy-ion
  collisions from $\sqrt{s}=200$ GeV to $\sqrt{s}=2.76$ TeV},''
  \href{http://dx.doi.org/10.1103/PhysRevC.84.014906}{{\em Phys. Rev.}
  {\bfseries C84} (2011) 014906},
\href{http://arxiv.org/abs/1103.5308}{{\ttfamily arXiv:1103.5308 [hep-ph]}}.
%%CITATION = ARXIV:1103.5308;%%.

\bibitem{E802Et:1992}
{\bfseries E802} Collaboration, T.~Abbott {\em et~al.}, ``{Global transverse
  energy distributions in relativistic nuclear collisions at 14.6A GeV/c},''
  {\em Phys. Rev.} {\bfseries C45} (1992) 2933--2952.

\bibitem{Lin:2000cx}
Z.-w. Lin, S.~Pal, C.~Ko, B.-A. Li, and B.~Zhang, ``{Charged particle rapidity
  distributions at relativistic energies},''
  \href{http://dx.doi.org/10.1103/PhysRevC.64.011902}{{\em Phys. Rev.}
  {\bfseries C64} (2001) 011902},
\href{http://arxiv.org/abs/nucl-th/0011059}{{\ttfamily arXiv:nucl-th/0011059
  [nucl-th]}}.
%%CITATION = NUCL-TH/0011059;%%.

\bibitem{Lokhtin:2005px}
I.~Lokhtin and A.~Snigirev, ``{A Model of jet quenching in ultrarelativistic
  heavy ion collisions and high-p(T) hadron spectra at RHIC},''
  \href{http://dx.doi.org/10.1140/epjc/s2005-02426-3}{{\em Eur. Phys. J.}
  {\bfseries C45} (2006) 211--217},
\href{http://arxiv.org/abs/hep-ph/0506189}{{\ttfamily arXiv:hep-ph/0506189
  [hep-ph]}}.
%%CITATION = HEP-PH/0506189;%%.

\bibitem{Albacete:2012xq}
J.~L. Albacete, A.~Dumitru, H.~Fujii, and Y.~Nara, ``{CGC predictions for p+Pb
  collisions at the LHC},''
  \href{http://dx.doi.org/10.1016/j.nuclphysa.2012.09.012}{{\em Nucl. Phys.}
  {\bfseries A897} (2013) 1--27},
\href{http://arxiv.org/abs/1209.2001}{{\ttfamily arXiv:1209.2001 [hep-ph]}}.
%%CITATION = ARXIV:1209.2001;%%.

\bibitem{Alver:2008zza}
B.~Alver {\em et~al.}, ``{Importance of correlations and fluctuations on the
  initial source eccentricity in high-energy nucleus-nucleus collisions},''
  \href{http://dx.doi.org/10.1103/PhysRevC.77.014906}{{\em Phys. Rev.}
  {\bfseries C77} (2008) 014906},
\href{http://arxiv.org/abs/0711.3724}{{\ttfamily arXiv:0711.3724 [nucl-ex]}}.
%%CITATION = ARXIV:0711.3724;%%.

\end{thebibliography}\endgroup
\newpage
\appendix
\section{The ALICE Collaboration}
\label{app:collab}

% Collaboration: CERN-LHC-ALICE
% Generation Date is 2016/Feb/19

% How to use:
%%%%%%%%% appendix with author list
%\appendix
%\section{The ALICE Collaboration}
%\label{app:collab}
%\input{authors-list.tex}  %%%%%%% get the latest version before submitting

\begingroup
\small
\begin{flushleft}
J.~Adam\Irefn{org39}\And
D.~Adamov\'{a}\Irefn{org84}\And
M.M.~Aggarwal\Irefn{org88}\And
G.~Aglieri Rinella\Irefn{org35}\And
M.~Agnello\Irefn{org110}\And
N.~Agrawal\Irefn{org47}\And
Z.~Ahammed\Irefn{org133}\And
S.~Ahmad\Irefn{org19}\And
S.U.~Ahn\Irefn{org68}\And
S.~Aiola\Irefn{org137}
% ...
\renewcommand\labelenumi{\textsuperscript{\theenumi}~}

\section*{Affiliation notes}
\renewcommand\theenumi{\roman{enumi}}
\begin{Authlist}
\item \Adef{0}Deceased
\item \Adef{idp1775680}{Also at: Georgia State University, Atlanta, Georgia, United States}
\item \Adef{idp3130944}{Also at: Also at Department of Applied Physics, Aligarh Muslim University, Aligarh, India}
\item \Adef{idp3839168}{Also at: M.V. Lomonosov Moscow State University, D.V. Skobeltsyn Institute of Nuclear, Physics, Moscow, Russia}
\end{Authlist}

\section*{Collaboration Institutes}
\renewcommand\theenumi{\arabic{enumi}~}
\begin{Authlist}

\item \Idef{org1}A.I. Alikhanyan National Science Laboratory (Yerevan Physics Institute) Foundation, Yerevan, Armenia
\item \Idef{org2}Benem\'{e}rita Universidad Aut\'{o}noma de Puebla, Puebla, Mexico
\item \Idef{org3}Bogolyubov Institute for Theoretical Physics, Kiev, Ukraine
\item \Idef{org4}Bose Institute, Department of Physics and Centre for Astroparticle Physics and Space Science (CAPSS), Kolkata, India
\item \Idef{org5}Budker Institute for Nuclear Physics, Novosibirsk, Russia
\item \Idef{org6}California Polytechnic State University, San Luis Obispo, California, United States
\item \Idef{org7}Central China Normal University, Wuhan, China
\item \Idef{org8}Centre de Calcul de l'IN2P3, Villeurbanne, France
\item \Idef{org9}Centro de Aplicaciones Tecnol\'{o}gicas y Desarrollo Nuclear (CEADEN), Havana, Cuba
\item \Idef{org10}Centro de Investigaciones Energ\'{e}ticas Medioambientales y Tecnol\'{o}gicas (CIEMAT), Madrid, Spain
\item \Idef{org11}Centro de Investigaci\'{o}n y de Estudios Avanzados (CINVESTAV), Mexico City and M\'{e}rida, Mexico
\item \Idef{org12}Centro Fermi - Museo Storico della Fisica e Centro Studi e Ricerche ``Enrico Fermi'', Rome, Italy
\item \Idef{org13}Chicago State University, Chicago, Illinois, USA
\item \Idef{org14}China Institute of Atomic Energy, Beijing, China
\item \Idef{org15}Commissariat \`{a} l'Energie Atomique, IRFU, Saclay, France
\item \Idef{org16}COMSATS Institute of Information Technology (CIIT), Islamabad, Pakistan
\item \Idef{org17}Departamento de F\'{\i}sica de Part\'{\i}culas and IGFAE, Universidad de Santiago de Compostela, Santiago de Compostela, Spain
\item \Idef{org18}Department of Physics and Technology, University of Bergen, Bergen, Norway
\item \Idef{org19}Department of Physics, Aligarh Muslim University, Aligarh, India
\item \Idef{org20}Department of Physics, Ohio State University, Columbus, Ohio, United States
\item \Idef{org21}Department of Physics, Sejong University, Seoul, South Korea
\item \Idef{org22}Department of Physics, University of Oslo, Oslo, Norway
\item \Idef{org23}Dipartimento di Fisica dell'Universit\`{a} 'La Sapienza' and Sezione INFN Rome, Italy
\item \Idef{org24}Dipartimento di Fisica dell'Universit\`{a} and Sezione INFN, Cagliari, Italy
\item \Idef{org25}Dipartimento di Fisica dell'Universit\`{a} and Sezione INFN, Trieste, Italy
\item \Idef{org26}Dipartimento di Fisica dell'Universit\`{a} and Sezione INFN, Turin, Italy
\item \Idef{org27}Dipartimento di Fisica e Astronomia dell'Universit\`{a} and Sezione INFN, Bologna, Italy
\item \Idef{org28}Dipartimento di Fisica e Astronomia dell'Universit\`{a} and Sezione INFN, Catania, Italy
\item \Idef{org29}Dipartimento di Fisica e Astronomia dell'Universit\`{a} and Sezione INFN, Padova, Italy
\item \Idef{org30}Dipartimento di Fisica `E.R.~Caianiello' dell'Universit\`{a} and Gruppo Collegato INFN, Salerno, Italy
\item \Idef{org31}Dipartimento di Scienze e Innovazione Tecnologica dell'Universit\`{a} del  Piemonte Orientale and Gruppo Collegato INFN, Alessandria, Italy
\item \Idef{org32}Dipartimento Interateneo di Fisica `M.~Merlin' and Sezione INFN, Bari, Italy
\item \Idef{org33}Division of Experimental High Energy Physics, University of Lund, Lund, Sweden
\item \Idef{org34}Eberhard Karls Universit\"{a}t T\"{u}bingen, T\"{u}bingen, Germany
\item \Idef{org35}European Organization for Nuclear Research (CERN), Geneva, Switzerland
\item \Idef{org36}Excellence Cluster Universe, Technische Universit\"{a}t M\"{u}nchen, Munich, Germany
\item \Idef{org37}Faculty of Engineering, Bergen University College, Bergen, Norway
\item \Idef{org38}Faculty of Mathematics, Physics and Informatics, Comenius University, Bratislava, Slovakia
\item \Idef{org39}Faculty of Nuclear Sciences and Physical Engineering, Czech Technical University in Prague, Prague, Czech Republic
\item \Idef{org40}Faculty of Science, P.J.~\v{S}af\'{a}rik University, Ko\v{s}ice, Slovakia
\item \Idef{org41}Faculty of Technology, Buskerud and Vestfold University College, Vestfold, Norway
\item \Idef{org42}Frankfurt Institute for Advanced Studies, Johann Wolfgang Goethe-Universit\"{a}t Frankfurt, Frankfurt, Germany
\item \Idef{org43}Gangneung-Wonju National University, Gangneung, South Korea
\item \Idef{org44}Gauhati University, Department of Physics, Guwahati, India
\item \Idef{org45}Helsinki Institute of Physics (HIP), Helsinki, Finland
\item \Idef{org46}Hiroshima University, Hiroshima, Japan
\item \Idef{org47}Indian Institute of Technology Bombay (IIT), Mumbai, India
\item \Idef{org48}Indian Institute of Technology Indore, Indore (IITI), India
\item \Idef{org49}Indonesian Institute of Sciences, Jakarta, Indonesia
\item \Idef{org50}Inha University, Incheon, South Korea
\item \Idef{org51}Institut de Physique Nucl\'eaire d'Orsay (IPNO), Universit\'e Paris-Sud, CNRS-IN2P3, Orsay, France
\item \Idef{org52}Institut f\"{u}r Informatik, Johann Wolfgang Goethe-Universit\"{a}t Frankfurt, Frankfurt, Germany
\item \Idef{org53}Institut f\"{u}r Kernphysik, Johann Wolfgang Goethe-Universit\"{a}t Frankfurt, Frankfurt, Germany
\item \Idef{org54}Institut f\"{u}r Kernphysik, Westf\"{a}lische Wilhelms-Universit\"{a}t M\"{u}nster, M\"{u}nster, Germany
\item \Idef{org55}Institut Pluridisciplinaire Hubert Curien (IPHC), Universit\'{e} de Strasbourg, CNRS-IN2P3, Strasbourg, France
\item \Idef{org56}Institute for Nuclear Research, Academy of Sciences, Moscow, Russia
\item \Idef{org57}Institute for Subatomic Physics of Utrecht University, Utrecht, Netherlands
\item \Idef{org58}Institute for Theoretical and Experimental Physics, Moscow, Russia
\item \Idef{org59}Institute of Experimental Physics, Slovak Academy of Sciences, Ko\v{s}ice, Slovakia
\item \Idef{org60}Institute of Physics, Academy of Sciences of the Czech Republic, Prague, Czech Republic
\item \Idef{org61}Institute of Physics, Bhubaneswar, India
\item \Idef{org62}Institute of Space Science (ISS), Bucharest, Romania
\item \Idef{org63}Instituto de Ciencias Nucleares, Universidad Nacional Aut\'{o}noma de M\'{e}xico, Mexico City, Mexico
\item \Idef{org64}Instituto de F\'{\i}sica, Universidad Nacional Aut\'{o}noma de M\'{e}xico, Mexico City, Mexico
\item \Idef{org65}iThemba LABS, National Research Foundation, Somerset West, South Africa
\item \Idef{org66}Joint Institute for Nuclear Research (JINR), Dubna, Russia
\item \Idef{org67}Konkuk University, Seoul, South Korea
\item \Idef{org68}Korea Institute of Science and Technology Information, Daejeon, South Korea
\item \Idef{org69}KTO Karatay University, Konya, Turkey
\item \Idef{org70}Laboratoire de Physique Corpusculaire (LPC), Clermont Universit\'{e}, Universit\'{e} Blaise Pascal, CNRS--IN2P3, Clermont-Ferrand, France
\item \Idef{org71}Laboratoire de Physique Subatomique et de Cosmologie, Universit\'{e} Grenoble-Alpes, CNRS-IN2P3, Grenoble, France
\item \Idef{org72}Laboratori Nazionali di Frascati, INFN, Frascati, Italy
\item \Idef{org73}Laboratori Nazionali di Legnaro, INFN, Legnaro, Italy
\item \Idef{org74}Lawrence Berkeley National Laboratory, Berkeley, California, United States
\item \Idef{org75}Moscow Engineering Physics Institute, Moscow, Russia
\item \Idef{org76}Nagasaki Institute of Applied Science, Nagasaki, Japan
\item \Idef{org77}National Centre for Nuclear Studies, Warsaw, Poland
\item \Idef{org78}National Institute for Physics and Nuclear Engineering, Bucharest, Romania
\item \Idef{org79}National Institute of Science Education and Research, Bhubaneswar, India
\item \Idef{org80}National Research Centre Kurchatov Institute, Moscow, Russia
\item \Idef{org81}Niels Bohr Institute, University of Copenhagen, Copenhagen, Denmark
\item \Idef{org82}Nikhef, Nationaal instituut voor subatomaire fysica, Amsterdam, Netherlands
\item \Idef{org83}Nuclear Physics Group, STFC Daresbury Laboratory, Daresbury, United Kingdom
\item \Idef{org84}Nuclear Physics Institute, Academy of Sciences of the Czech Republic, \v{R}e\v{z} u Prahy, Czech Republic
\item \Idef{org85}Oak Ridge National Laboratory, Oak Ridge, Tennessee, United States
\item \Idef{org86}Petersburg Nuclear Physics Institute, Gatchina, Russia
\item \Idef{org87}Physics Department, Creighton University, Omaha, Nebraska, United States
\item \Idef{org88}Physics Department, Panjab University, Chandigarh, India
\item \Idef{org89}Physics Department, University of Athens, Athens, Greece
\item \Idef{org90}Physics Department, University of Cape Town, Cape Town, South Africa
\item \Idef{org91}Physics Department, University of Jammu, Jammu, India
\item \Idef{org92}Physics Department, University of Rajasthan, Jaipur, India
\item \Idef{org93}Physik Department, Technische Universit\"{a}t M\"{u}nchen, Munich, Germany
\item \Idef{org94}Physikalisches Institut, Ruprecht-Karls-Universit\"{a}t Heidelberg, Heidelberg, Germany
\item \Idef{org95}Purdue University, West Lafayette, Indiana, United States
\item \Idef{org96}Pusan National University, Pusan, South Korea
\item \Idef{org97}Research Division and ExtreMe Matter Institute EMMI, GSI Helmholtzzentrum f\"ur Schwerionenforschung, Darmstadt, Germany
\item \Idef{org98}Rudjer Bo\v{s}kovi\'{c} Institute, Zagreb, Croatia
\item \Idef{org99}Russian Federal Nuclear Center (VNIIEF), Sarov, Russia
\item \Idef{org100}Saha Institute of Nuclear Physics, Kolkata, India
\item \Idef{org101}School of Physics and Astronomy, University of Birmingham, Birmingham, United Kingdom
\item \Idef{org102}Secci\'{o}n F\'{\i}sica, Departamento de Ciencias, Pontificia Universidad Cat\'{o}lica del Per\'{u}, Lima, Peru
\item \Idef{org103}Sezione INFN, Bari, Italy
\item \Idef{org104}Sezione INFN, Bologna, Italy
\item \Idef{org105}Sezione INFN, Cagliari, Italy
\item \Idef{org106}Sezione INFN, Catania, Italy
\item \Idef{org107}Sezione INFN, Padova, Italy
\item \Idef{org108}Sezione INFN, Rome, Italy
\item \Idef{org109}Sezione INFN, Trieste, Italy
\item \Idef{org110}Sezione INFN, Turin, Italy
\item \Idef{org111}SSC IHEP of NRC Kurchatov institute, Protvino, Russia
\item \Idef{org112}Stefan Meyer Institut f\"{u}r Subatomare Physik (SMI), Vienna, Austria
\item \Idef{org113}SUBATECH, Ecole des Mines de Nantes, Universit\'{e} de Nantes, CNRS-IN2P3, Nantes, France
\item \Idef{org114}Suranaree University of Technology, Nakhon Ratchasima, Thailand
\item \Idef{org115}Technical University of Ko\v{s}ice, Ko\v{s}ice, Slovakia
\item \Idef{org116}Technical University of Split FESB, Split, Croatia
\item \Idef{org117}The Henryk Niewodniczanski Institute of Nuclear Physics, Polish Academy of Sciences, Cracow, Poland
\item \Idef{org118}The University of Texas at Austin, Physics Department, Austin, Texas, USA
\item \Idef{org119}Universidad Aut\'{o}noma de Sinaloa, Culiac\'{a}n, Mexico
\item \Idef{org120}Universidade de S\~{a}o Paulo (USP), S\~{a}o Paulo, Brazil
\item \Idef{org121}Universidade Estadual de Campinas (UNICAMP), Campinas, Brazil
\item \Idef{org122}University of Houston, Houston, Texas, United States
\item \Idef{org123}University of Jyv\"{a}skyl\"{a}, Jyv\"{a}skyl\"{a}, Finland
\item \Idef{org124}University of Liverpool, Liverpool, United Kingdom
\item \Idef{org125}University of Tennessee, Knoxville, Tennessee, United States
\item \Idef{org126}University of the Witwatersrand, Johannesburg, South Africa
\item \Idef{org127}University of Tokyo, Tokyo, Japan
\item \Idef{org128}University of Tsukuba, Tsukuba, Japan
\item \Idef{org129}University of Zagreb, Zagreb, Croatia
\item \Idef{org130}Universit\'{e} de Lyon, Universit\'{e} Lyon 1, CNRS/IN2P3, IPN-Lyon, Villeurbanne, France
\item \Idef{org131}Universit\`{a} di Brescia
\item \Idef{org132}V.~Fock Institute for Physics, St. Petersburg State University, St. Petersburg, Russia
\item \Idef{org133}Variable Energy Cyclotron Centre, Kolkata, India
\item \Idef{org134}Warsaw University of Technology, Warsaw, Poland
\item \Idef{org135}Wayne State University, Detroit, Michigan, United States
\item \Idef{org136}Wigner Research Centre for Physics, Hungarian Academy of Sciences, Budapest, Hungary
\item \Idef{org137}Yale University, New Haven, Connecticut, United States
\item \Idef{org138}Yonsei University, Seoul, South Korea
\item \Idef{org139}Zentrum f\"{u}r Technologietransfer und Telekommunikation (ZTT), Fachhochschule Worms, Worms, Germany
\end{Authlist}
\endgroup

  %%%%%%% done by webmaster team
\end{document}